\documentclass[10pt,twocolumn,floatfix,prb,superscriptaddress]{revtex4-2}

\usepackage{amsmath,amssymb,amsthm,mathrsfs,amsfonts,dsfont,amstext} 
\usepackage{textcomp,pbox}
\usepackage[export]{adjustbox}
\usepackage{bm}
\usepackage{dcolumn,booktabs,url}
\usepackage[scaled]{helvet}
\usepackage{sansmath,gensymb}
\usepackage{tikz,graphicx,transparent,color}
\usepackage{multirow}
\usepackage[separate-uncertainty = true]{siunitx}
\usepackage{comment}
\usepackage{physics}
\usepackage{pdfpages}
\usepackage{array}
\usepackage{makecell}
\usepackage{tabularx}
\usepackage{dsfont}

\newcolumntype{L}{>{\raggedright\arraybackslash}X}

\makeatletter
\AtBeginDocument{\let\LS@rot\@undefined}
\makeatother

\newcolumntype{C}[1]{>{\centering\let\newline\\\arraybackslash\hspace{0pt}}m{#1}}

\usepackage[colorlinks=true]{hyperref}
\usepackage{graphicx}

\hypersetup{
     colorlinks   = true,
     citecolor    = red,
     linkcolor    = blue,
     urlcolor     = red     
}

\graphicspath{{./figs/}}

\newcommand{\figref}[1]{\figurename{~\ref{#1}}}
\newcommand{\tabref}[1]{\tablename{~\ref{#1}}}

\usepackage{lipsum}% http://ctan.org/pkg/lipsum
\usepackage{graphicx}% http://ctan.org/pkg/graphicx

\usepackage{import}
\usepackage{xifthen}
\usepackage{pdfpages}
\usepackage{transparent}
\usepackage{svg}
\usepackage{adjustbox}
\usepackage{calc}

\usepackage{physics}

\makeatletter
\def\maketitle{
\@author@finish
\title@column\titleblock@produce
\suppressfloats[t]}
\makeatother

\usepackage[resetlabels]{multibib}

\newcites{S}{References}

\begin{document}

\newcommand{\TitleName}{Collective super- and subradiant dynamics between distant optical quantum emitters\\
} 

\title{\TitleName}

\newcommand{\AffCPH}{Center for Hybrid Quantum Networks (Hy-Q), The Niels Bohr Institute, University~of~Copenhagen,  DK-2100  Copenhagen~{\O}, Denmark}
\newcommand{\AffBochum}{Lehrstuhl f\"ur Angewandte Festk\"orperphysik, Ruhr-Universit\"at Bochum, Universit\"atsstra\ss e 150, D-44801 Bochum, Germany}

\author{Alexey Tiranov$^{\dag}$}
\thanks{These authors contributed equally to this work.}
\affiliation{\AffCPH{}}

\author{Vasiliki Angelopoulou}
\thanks{These authors contributed equally to this work.}
\affiliation{\AffCPH{}}

\author{Cornelis Jacobus van Diepen}
\thanks{These authors contributed equally to this work.}
\affiliation{\AffCPH{}}

\author{Bj\"{o}rn Schrinski}
\affiliation{\AffCPH{}}
\author{Oliver August Dall'Alba Sandberg}
\affiliation{\AffCPH{}}
\author{Ying Wang}
\affiliation{\AffCPH{}}
\author{Leonardo Midolo}
\affiliation{\AffCPH{}}
\author{Sven Scholz}
\affiliation{\AffBochum{}}
\author{Andreas Dirk Wieck}
\affiliation{\AffBochum{}}
\author{Arne Ludwig}
\affiliation{\AffBochum{}}
\author{Anders S{\o}ndberg S{\o}rensen}
\affiliation{\AffCPH{}}
\author{Peter Lodahl}
\thanks{Email to: alexey.tiranov@nbi.ku.dk; lodahl@nbi.ku.dk}
\affiliation{\AffCPH{}}

\date{\today}

\begin{abstract}
Photon emission is the hallmark of light-matter interaction and the foundation of photonic quantum science, enabling advanced  sources for quantum communication and computing. While single-emitter radiation can be tailored by the photonic environment, the introduction of multiple emitters extends this picture. A fundamental challenge, however, is that the radiative dipole-dipole coupling rapidly decays with spatial separation, typically within a fraction of the optical wavelength. We realize distant dipole-dipole radiative coupling with pairs of solid-state optical quantum emitters embedded in a nanophotonic waveguide. We dynamically probe the collective response and identify both super- and subradiant emission as well as means to control the dynamics by proper excitation techniques. Our work constitutes a foundational step towards multi-emitter applications for scalable quantum information processing. 
\end{abstract}

\maketitle 

The radiative coupling of multiple optical emitters has been a long-standing challenge in quantum optics and atomic physics~\cite{Dicke1954,GROSS1982,DeVoe1996,Eschner2001}. It offers a route to realizing quantum gates between emitters \cite{Beige2000} thereby constituting a fundamental building block for quantum-information processing \cite{OBrien2009}. Waveguide quantum electrodynamics (QED) \cite{Chang2018}
has evolved as a research discipline ideally suited for overcoming the inherently weak dipole-dipole coupling. This is  because the radiative coupling here is extended significantly beyond the sub-wavelength limit encountered in unstructured media \cite{Trebbia2022}. The dipole-dipole interaction can be understood as the absorption and re-emission of virtual photons, and the waveguide extends the spatial range to be limited only by the weak leakage of the waveguide mode due to structural imperfections \cite{Lodahl2015}.  
It leads to the formation of collective emitter states featuring super- or subradiant decay rates \cite{Dicke1954,GROSS1982,Scully2015}, as controlled by the optical phase lag between the two emitters (\figref{fig:1}a).  
Observation of collective emission requires a highly coherent light-matter interface. In particular, long-lived subradiant features may be elusive in the presence of experimental imperfections such as dephasing \cite{Chu2022}.

Collective multi-emitter effects have been studied in the optical~\cite{DeVoe1996,Eschner2001,Goban2015,Sipahigil2016,Solano2017,Evans2018,Trebbia2022} and microwave domains ~\cite{Loo2013,Mlynek2014,Mirhosseini2019}. In microwave QED, multi-qubit interactions have been realized ~\cite{Brehm2021,Kim2021} and subradiant collective states coherently controlled~\cite{Zanner2022}. Realizing such functionalities in the optical domain is essential: optical photons can be highly integrated, rapidly processed on-chip, and transmitted over extended distances \cite{OBrien2009,Uppu2021}, making photonics the backbone technology for the quantum internet~\cite{Kimble2008}. Previous reports in the optical domain include projective preparation of single-excitation superradiant states as revealed in two-photon correlation measurements \cite{Kim2018,Grim2019,Koong2021,Grim2022} that does not require emitter-emitter coupling. Spectroscopic evidence for coherent coupling was reported with closely spaced dye molecules in a bulk medium \cite{Hettich2002,Trebbia2022} and for vacancy centers in a cavity \cite{Evans2018}. Dynamics with modified collective emission would constitute direct experimental proof of coherent coupling and open new avenues towards applications. Here we report the  observation of coherent dynamics of the collective excitation of quantum dot (QD) emitters coupled via a photonic crystal waveguide (PCW) (\figref{fig:1}a) by observing both enhanced (superradiant) and suppressed (subradiant) dynamics. 
In contrast to cooperative \cite{Lehmberg1970} and amplified \cite{Allen1973} spontaneous emission involving multiple excitations, here we concentrate on a single excitation distributed between a pair of QDs. In the former case, the resulting emission peak intensity scales as $N^2$, while in the latter, the enhancement is $\propto N$, where $N$ is the number of emitters involved. In the experiment, the QDs are brought into mutual resonance by Zeeman-tuning with a magnetic field. The coherent oscillations of the collective state are directly observed and found to be controllable by spectrally tuning the QDs and by varying the excitation conditions.

\begin{figure}[h!]
	\begin{center}
	\includegraphics[width=1\linewidth]{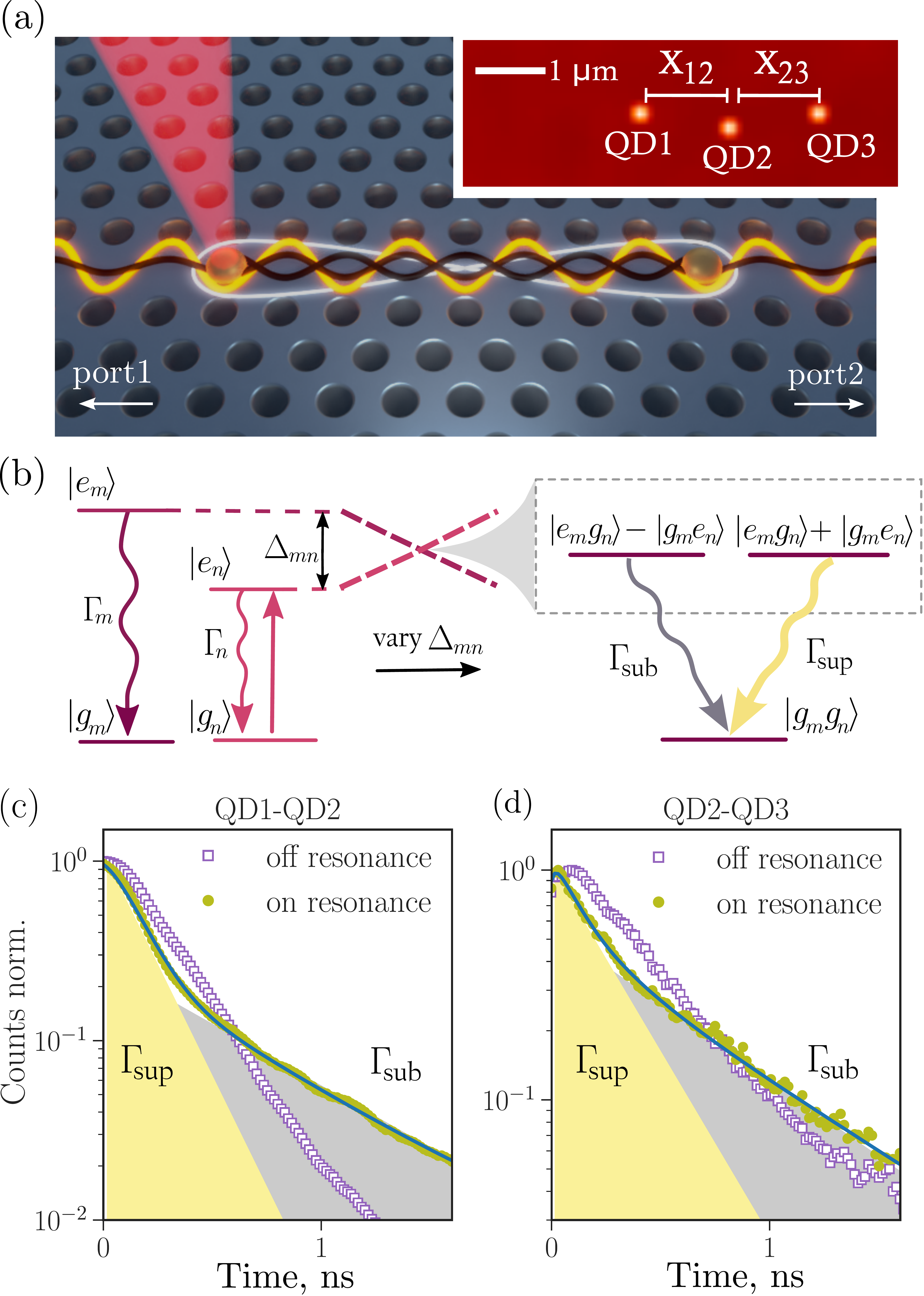}
	\end{center}
	\caption{ (color online) \textbf{Observation of the super- and subradiant emission.}
	(a) Illustration of the photon-mediated coupling between two QDs in a PCW where one QD is optically pumped. Subsequently, the emission dynamics of the coupled QD system exhibit super- and subradiance originating from either constructive (bright line) or destructive (dark line) interference of the field emitted into the PCW and scattered by each of the QDs.
	The inset shows a fluorescence image of the QDs in the PCW where  $x_{12} = 1.25(3)$~\textmu m, $x_{23} = 0.96(3)$~\textmu m.
	(b) Energy level diagram of two QDs with a mutual detuning of $\Delta_{mn}$. Off resonance (left image), the two QDs decay independently of each other, while on resonance (right image), super- and subradiant dynamics occur with rates $\Gamma_{\text{sup}}$ and $\Gamma_{\text{sub}}$, respectively.
	(c) and (d)	Measured time-resolved emission dynamics of pair QD1-QD2 and QD2-QD3 both on and off resonance and after exciting QD2 and QD3 through higher energy excited states. The measured count rates are normalized to the maximum at zero time delay corresponding to the time of excitation of the coupled system. The solid lines are the model fits to the experimental data. Off resonance, $\Delta_{12}/2\pi = 6$~GHz, $(\Delta_{23}/2\pi = 5$~GHz) the decay curve corresponds to the cases of individual QDs.  When two QDs are brought into resonance, two decay components are observed in the data, a fast $\Gamma_{\text{sup}}$ and a slow $\Gamma_{\text{sub}}$  corresponding to super- and subradiant dynamics. 
}
	\label{fig:1}
\end{figure}

\textbf{Hamiltonian for waveguide-mediated interaction.}

The coupled QDs are described by the  effective Hamiltonian $\mathcal{H}_\text{eff}$ (frame rotating at the excitation field frequency and  $\hbar=1$) \cite{Asenjo2017,Albrecht2019}
\begin{equation}\label{eq:EffHamiltonian}
     \mathcal{H}_\text{eff} = \sum_{m,n=1}^2 \left( J_{mn} - i \frac{\Gamma_{mn}}{2}\right) \sigma_n^+\sigma_m^{-} + \sum_{n=1}^{2} \left(\Delta_n - i \frac{\gamma_n^s}{2} \right)  \sigma_{n}^+\sigma^{-}_{n},
\end{equation}
where $J_{mn} = \frac{1}{2}\sqrt{\beta_m \beta_n \Gamma_m\Gamma_n} \sin\phi_{mn}$ and $\Gamma_{mn} = \sqrt{\beta_m \beta_n \Gamma_m \Gamma_n} \cos\phi_{mn}$ are the dispersive and dissipative coupling rates connecting QD$_m$ and QD$_n$.   $\Gamma_m = \gamma^{wg}_m + \gamma^s_m$  contains the decay rate into (out of) the waveguide $\gamma^{wg}_m$  $(\gamma_m^{s})$, corresponding to $\beta$-factors  $\beta_m = \gamma^{wg}_m/\Gamma_m$. $\Delta_m$ is the detuning of  QD$_m$ with respect to the excitation field frequency, such that $\Delta_{mn} = \Delta_m - \Delta_n$  is the detuning between the two QDs, $\phi_{mn} = k\abs{x_{mn}}$ is the phase lag due to the emitter separation $x_{mn}$ with $k$ being the effective wavenumber of the PCW mode. $\phi_{mn}$ determines the character of the coupling between dispersive ($\Gamma_{mn}=0$), which modifies the energy levels, to dissipative ($J_{mn}=0$), which affects the decay dynamics. The system bears a resemblance to the case of a nanocavity where each QD acts as an end mirror by scattering single photons into the mode of the waveguide, and $\phi_{mn}$ determines the cavity resonance condition (\figref{fig:1}a). Notably, the system is inherent in quantum character since the QDs only scatter a single photon at a time. $\sigma_m^+,\sigma_n^-$ are the raising and lowering operators for the optical transition of QD$_m$, QD$_n.$  

By selectively pumping  one of the QDs, e.g QD$_n$, one excitation is launched to  populate the state $\ket{g_me_n}$. On resonance $(\Delta_{mn}=0)$, the subsequent dynamics is best described in terms of the entangled eigenstates $\ket{S} = (\ket{e_mg_n}+\ket{g_me_n})/\sqrt{2}$ and $\ket{s} = (\ket{e_mg_n}-\ket{g_me_n})/\sqrt{2}$ with the associated decay rates, determined by the phase lag $\phi_{mn}$. The case when $\phi_{mn} = N \pi$ ($N$ integer) corresponds to a dissipative coupling between QDs leading to a superradiant $\ket{S}$ and a subradiant $\ket{s}$ state with modified decay rates (\figref{fig:1}b). $\phi_{mn} = (N+1/2)\pi$ results in dispersive coupling where the collective states are shifted by $J_{mn}$ while leaving their decay rates unchanged compared to the uncoupled emitters. In the present experiment, we are primarily studying the regime of dissipative radiative coupling, leading to modified emission dynamics.

\begin{figure*}
	\includegraphics[width=0.99\linewidth, trim=0.0cm .0cm 0.0cm 0.0cm,clip]{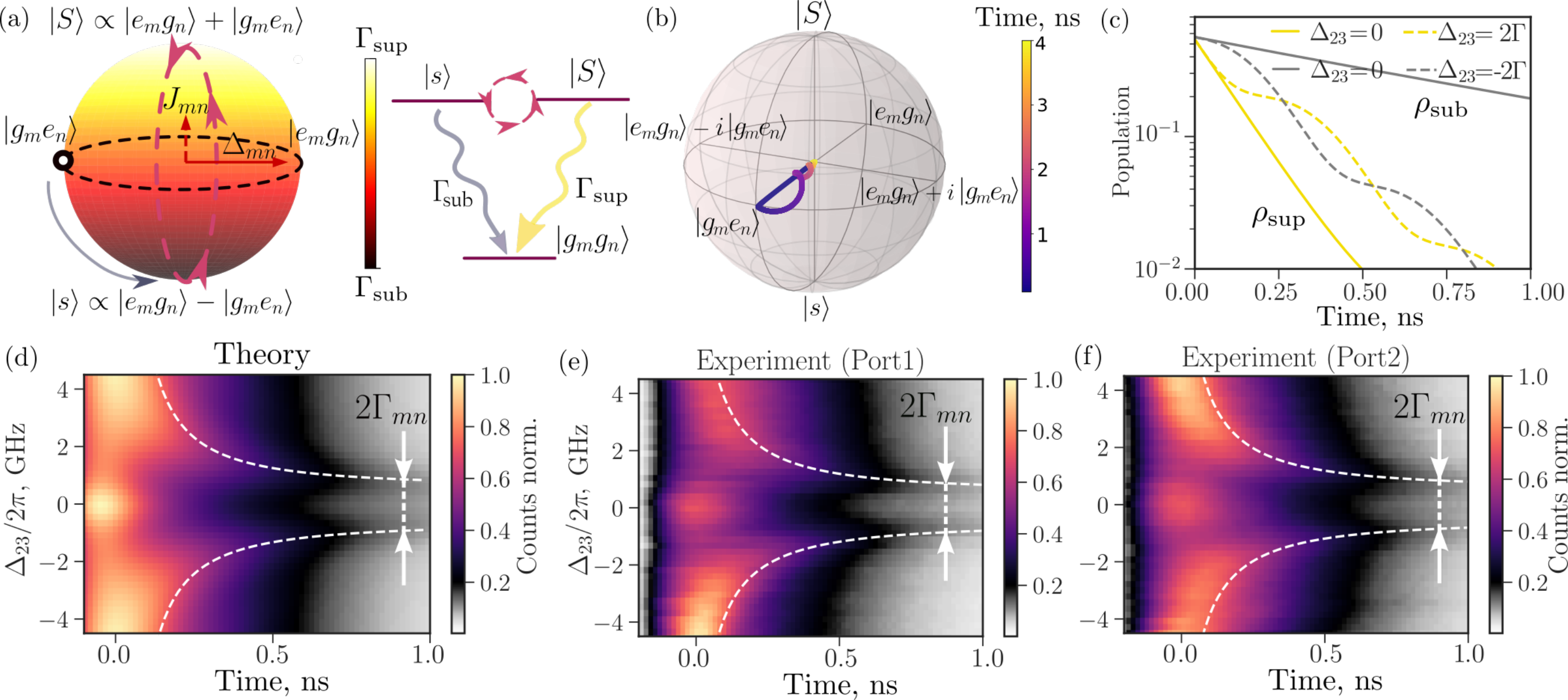}
	\caption{ (color online) 
 \textbf{Coherent dynamics of  coupled QDs}. 
	(a) The evolution of a single collective excitation can be represented on a Bloch sphere, with the color indicating the decay rate of the respective collective state. After exciting QD$_n$, the state  $\ket{g_me_n}$ is populated. Two processes occur: exponential decay by spontaneous emission (black solid arrow) and coherent evolution between super- and subradiant states (red arrow) as determined by the detuning between the two QDs $\Delta_{mn}$. (b) Exemplary   state evolution trajectory for $\Delta_{23} = -5.5\Gamma$, where the length of the Bloch vector shows the population in the single-excitation subspace. The color bar tracks the evolution time from preparing the coupled system in the initial state  until decaying to the ground state $\ket{g_mg_n}$ (origin of Bloch sphere).
	(c) Calculated population of the collective states as a function of time for two values of detuning and comparing super- (yellow curves) and subradiant (grey curves) contributions for an initial state $\ket{g_me_n}$.
	The corresponding emission intensity is plotted in (d).  The calculation is done with the experimentally extracted parameters of the QD2-QD3 pair (see Supplemental Notes).
	(e) Full experimental dataset for a continuous scan of detuning $\Delta_{23}$ and comparing measurements from outcoupling port 1 (e) and port 2 (f).  The dashed lines trace the maximum population after the first oscillation  and are given by $t = \pi/f_{\text{osc}}$, see main text for the definitions.  
}
	\label{fig:2}
\end{figure*}

\textbf{Super- and subradiance with coherent evolution.}
We optically excite a single QD   and record the collective emission dynamics from either collection port 1 or 2, studying three different QDs (QD1-3, see inset of \figref{fig:1}a). Two examples of recorded emission signals are shown on \figref{fig:1}c,d, where we alter the detuning to compare the off- and on-resonant cases; see also Supplemental Notes for further experimental details~\cite{SMref}. On resonance, we observe strongly modified decay dynamics due to coherent coupling, and both super- and subradiant features are directly visible. For QD1-QD2, we find radiative linewidths of $\Gamma_{\text{sup}}/2\pi = 1.36(8)$~GHz and $\Gamma_{\text{sub}}/2\pi = 0.280(2)$ GHz, by modeling the data with a bi-exponential decay.
The modeling of the data at short time delays is limited by the finite instrument response function of the single-photon detectors, particularly visible at the rising edge of the detected pulse.
These values should be referred to the single emitter linewidths of $\Gamma_{\text{2}}/2\pi = 0.79(2)$~GHz and $\Gamma_{\text{1}}/2\pi = 0.85(1)$~GHz, respectively, as  observed far off-resonance where the coupling is negligible. We derive a super/subradiant enhancement factor of $1.36/0.28 = 4.8$. The enhancement factor is a direct figure-of-merit of the collective coupling quality and is highly sensitive to experimental imperfections and decoherence; for a detailed account of the underlying physical parameters, see Supplemental Notes Table~2~\cite{SMref}. The long-range nature of the dipole-dipole coupling is explicitly demonstrated. Using resonant excitation through the PCW, we image the spatial separation of the QDs and find $x_{12} = 1.25(3)$~\textmu m and $x_{23}= 0.96(3)$~\textmu m (see inset of \figref{fig:1}a), which should be compared to the wavelength of $\lambda = 270$~nm inside GaAs or the PCW lattice constant of $a = 240$~nm.

\begin{figure*}
	\includegraphics[width=1\linewidth, trim=0.0cm .0cm 0.0cm 0.0cm,clip]{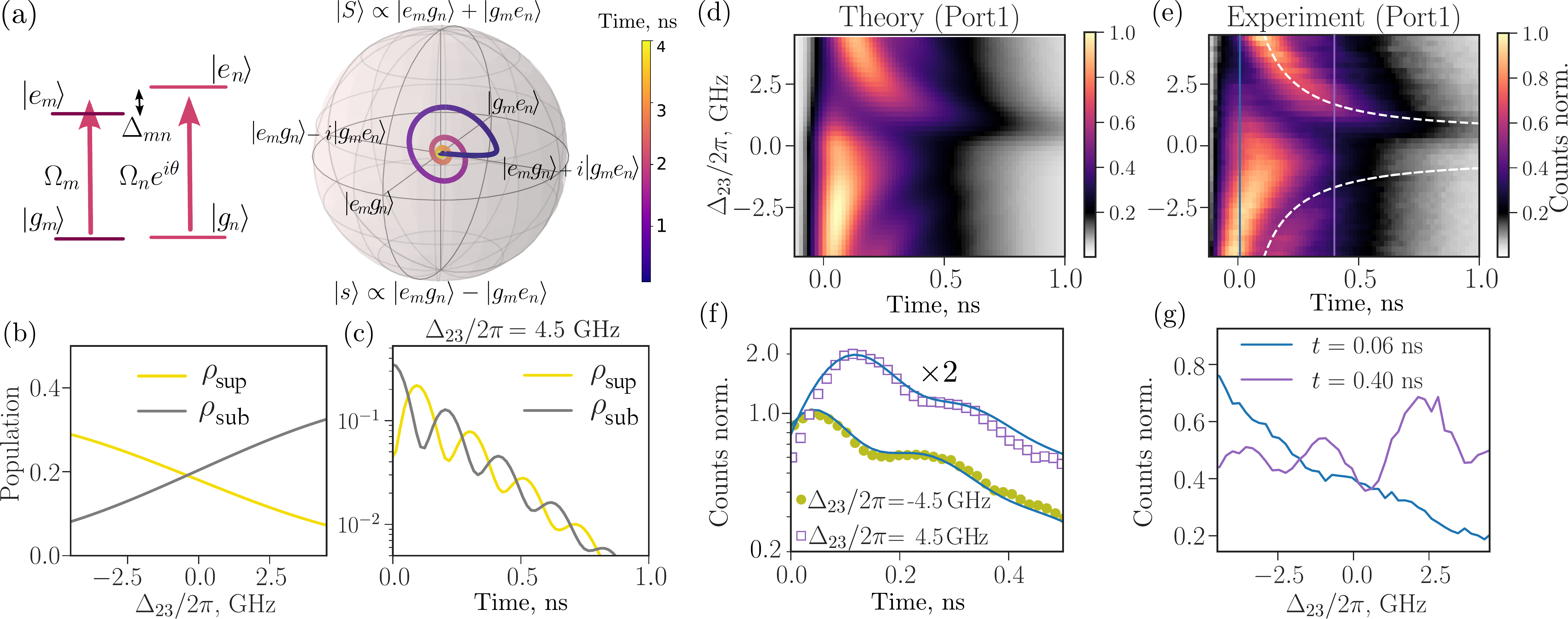}
	\caption{ 
    (color online) \textbf{Controlled preparation of the collective state}. 
	(a) Simultaneous driving of both QDs from QD2-QD3 pair with control of the phase $\theta$ allows to initialize the system in a collective state. In the case of $\theta \approx -\pi/2$ the initial state in the single-excitation subspace is close to $\ket{ge}+i\ket{eg}$. The detuning between the QDs results in the evolution of the collective state towards  $\ket{S}$ when $\Delta_{23}<0$ and towards $\ket{s}$ for $\Delta_{23}>0$. An example of the state trajectory for $\Delta_{23}<0$ is depicted on the Bloch sphere.
	(b) The  population of the super- $\rho_{\text{sup}}$ and subradiant $\rho_{\text{sub}}$ components during the excitation as a function of the detuning $\Delta_{23}$  for $\theta \approx -\pi/2$. The sign of the detuning determines whether the super- or subradiant states are preferentially populated. (c) The two populations oscillate out of phase as a function of time after excitation.
    (d) Calculated and (e) measured  intensity decay dynamics as a function of detuning and  collected from the outcoupling port 1. The asymmetric dependence on detuning  is well reproduced by the theory using parameters extracted from the experiment with $\theta \approx -\pi/2$.
	(f) Examples of emission intensity decay curves for two values of detuning explicitly displaying the coherent oscillations being out-of-phase of each other as determined by the sign of the detuning.  
    (g) Normalised counts as a function of  detuning and for two representative times quantifying the asymmetry.
}
	\label{fig:3}
\end{figure*}

Next, we show the coherent evolution of collective states by precisely controlling the detuning between two emitters. We start by resonantly exciting a single QD to prepare  $\ket{eg}=\left(\ket{S}+\ket{s}\right)/\sqrt{2}$, i.e., an equal superposition of the super- and subradiant states. The state subsequently evolves in time into a collective state (see the Bloch sphere graphical representation in \figref{fig:2}a,b). The color of the Bloch sphere surface represents the decay time of the respective state, and the collective state vector precesses due to coherent evolution. Assuming $\Gamma_m=\Gamma_n$, the correlated dynamics is described by the difference between the two eigenvalues  of the coupled system $f_{\text{osc}} = \mathrm{}\sqrt{\Delta_{mn}^2+\left(2J_{mn}-i\Gamma_{mn}\right)^2}$. E.g., in the case of pure dissipative coupling  ($J_{mn}=0$), two different regimes are identified (\figref{fig:2}c,d). In the underdamped regime, $\left| \Delta_{mn} \right| > \Gamma_{mn}$,  coherent evolution prevails over dissipation resulting in the observation of an increase in emission intensity. The dashed lines in \figref{fig:2}d-f track the observed intensity maximum for different values of  $\Delta_{mn}$ defined by $t = \pi/f_{\text{osc}}$. In the overdamped regime, $\left| \Delta_{mn} \right| < \Gamma_{mn}$, the dissipative coupling damps the excitation to the ground state faster than the coherent oscillations on the Bloch sphere (leading to the ``gap'' between two dashed lines on \figref{fig:2}d-f). $\Delta_{mn} = \Gamma_{mn}$ is the case of critical damping where the dissipation and coherent oscillation rates are balanced. These examples quantify how precise emitter tuning provides an experimentally accessible  ``control knob'' of coupled collective quantum states.

The rich coherent dynamics of the collectively-coupled system is evident from the experimental data (\figref{fig:2}e,f) that are well reproduced by theory (\figref{fig:2}d). As for the resonant case, following a fast decay of the superradiant component, the coupled system evolves toward the subradiant state. In the detuned case, coherent evolution is observed, leading to modified dynamics since super- and subradiant components are interchanged. The oscillation between super- and subradiant components, as defined by the detuning, leads in an increase of the intensity, corresponding to the superradiant component being maximal. This is clearly visible as a bright region, as it is tracked by the dashed line on \figref{fig:2}d-f.  By comparing the emission from two different directions (outcoupling port 1 and 2), we observe a similar behavior (\figref{fig:2}e,f), which is consistent with a predominantly dissipative coupling $\phi_{mn} \simeq N\pi$. The experiment was repeated on in total three pairs of QDs,  where the additional data and detailed modeling can be found in the Supplemental Notes~\cite{SMref}. For all three data sets, we observe predominantly dissipative coupling, which likely results from the fact that QD candidates, featuring efficient coupling to the PCW, were pre-selected in the experiment based on resonant transmission measurements through the waveguide (see Supplemental Notes~\cite{SMref}). This condition results in the selection of QDs close to the waveguide center and, therefore, a phase lag between QDs primarily determined by the  periodicity of the photonic crystal lattice. Since the PCW coupling is large and QDs are spectrally close to the band edge of the waveguide mode, this leads to $k/a \simeq \pi$~\cite{Lodahl2015}, whereby the phase lag between neighboring PCW unit cells is $\simeq \pi$, see Supplemental Notes for further details~\cite{SMref}. 

The theoretical model (\figref{fig:2}d) reproduces the experimental data  and fully captures the complex coherent quantum dynamics observed experimentally  (\figref{fig:2}e,f). From the analysis, we extract for pair QD2-QD3: $\Gamma_{23}/2\pi= 0.61$~GHz and $J_{23}/2\pi= 0.03$~GHz, respectively. The dissipative coupling rate is comparable to the intrinsic linewidth of the respective QDs, while the dispersive part is almost vanishing. 
In the case of negligible dispersive coupling the superradiant $\Gamma_{sup}$ (subradiant $\Gamma_{sub}$) decay rate can be approximated by $\Gamma_{sup} \approx \Gamma + \Gamma_{ij} - \sigma_{sd}^2/2\beta \Gamma$ ($\Gamma_{sub} \approx \Gamma-\Gamma_{ij} + \sigma_{sd}^2/2\beta \Gamma$), where $\sigma_{sd}$ is the spectral diffusion width. The predicted value of $\Gamma_{sup}/2\pi=1.25$~GHz ($\Gamma_{sub}/2\pi=0.27$~GHz) for the QD2-QD3 pair is in agreement with the experimentally measured $\Gamma_{sup}/2\pi=1.33$~GHz ($\Gamma_{sub}/2\pi=0.22$~GHz). For this QD pair, we obtain $\beta_2 = 0.88$ and $\beta_3=0.83$ from the resonant transmission data, see Supplementary Materials \cite{SMref} for further details.

As opposed to the case of superconducting qubits~\cite{Loo2013}, changing the detuning between the QDs has a negligible effect on the phase lag, whereby the coupling remains dissipative.  This is a consequence of the fact that the detuning is vanishingly small compared to the optical frequency. This distinction may be advantageous in applications of radiative collective coupling, since it allows the detuning to be exploited as a control parameter without changing the coupling, which is explicitly demonstrated in the present work. Interestingly, even when the emitter-emitter system is initialized in $\ket{eg}$ or $\ket{ge}$, coherent oscillations are still observed. This is enabled by the fast decay of the superradiant component after initialization, leading to the population of the slower decaying subradiant state that coherently evolves on the Bloch sphere (\figref{fig:2}b).

\textbf{Control of collective excitations.}
The deterministic preparation of collective states is essential in order to pave the way for their applications in  quantum-information processing. To this end, we coherently excite  both QDs in order to control the initial collective state on the Bloch sphere (\figref{fig:3}a). In the magnetic Zeeman field, the coupled QD transitions are orthogonally (circularly) polarized \cite{Warburton2013}, yet they are efficiently coupled by the optical mode of the PCW. The phase between the two driving fields $\Omega_2$ and $\Omega_3 e^{i\theta}$ is adjusted via the polarization of the excitation laser. Using a single laser, we implement $\theta \approx -\pi/2$, which prepares an initial state with the single excitation close to $\ket{eg}+i\ket{ge}$ for $\Delta_{23}=0$. With detuning, the state either evolves towards the super-  (for $\Delta_{23} <0$)  or the subradiant (for $\Delta_{23} >0$) state. This results in a striking difference in the radiation at a short time, followed by out-of-phase coherent oscillations between the two components (\figref{fig:3}b,c. 

The experimental demonstration of this behavior (\figref{fig:3}e) is accurately described by the theory (\figref{fig:3}d). We observe  a pronounced asymmetry around zero detuning, where for positive detunings, the emission dynamics is effectively delayed. This stems from the selective population of the  subradiant state resulting in a lower emission at early times. Subsequently, the emission intensity increases as the coherent evolution increases the population of the superradiant state. The reverse behavior is found for $\Delta_{23}<0$.  This is a result of the out-of-phase oscillation of the population of the super- and subradiant components (\figref{fig:3}f,g). As opposed to the case where a single QD was excited (\figref{fig:2}), we observe here multiple coherent oscillations. This is due to the state $\ket{eg}+i\ket{ge}$, starting to coherently evolve on the Bloch sphere directly after excitation. This is in contrast to the case where  $\ket{eg}$ is prepared, and the coherent evolution sets in once  the state has partially decayed to $\ket{s}$. We track the maximum emission intensity (dashed line) for $\Delta_{23} > 0$ (and minimum for $\Delta_{23} < 0$) in the plots of \figref{fig:3}e. It is consequently found that the collective light emission intensity can be controlled via the QD detuning. By simultaneous driving, we populate the doubly excited state $\ket{ee}$ component, which reaches 0.07 for $\approx\pi/3$ excitation pulses that are used. We note that the doubly excited state $\ket{ee}$, however, does not contribute to the measured asymmetry between positive and negative detunings.

\vspace{0.1cm}
\textbf{Concluding remarks.}
Our observation of super- and subradiant emission dynamics using pairs of QDs embedded in PCWs and separated by a distance much larger than the wavelength is facilitated by the PCW offering broadband spectral operation and long-range photon-mediated coherent interaction between QDs. Our work can constitute a foundational step towards multi-emitter applications of technological importance, e.g., for realizing quantum transduction between microwave qubits and the optical domain \cite{Elfving2019} or for quantum memories with exponential improvement in photon storage fidelity \cite{Asenjo2017}. The ability to manipulate the super- and subradiant state dynamics by controlling the detuning and pumping conditions will lead to a whole new range of opportunities when implementing a coherent spin inside the QD \cite{Warburton2013,Appel2021}. For instance, advanced photonic cluster states may be generated deterministically \cite{Economou2010}, providing a universal resource for measurement-based photonic quantum computing. To this end,  waveguide-mediated dissipative coupling can be exploited to realize spin-spin entanglement  between distant quantum emitters~\cite{Gullans2012}. Another direction will exploit photon scattering from coupled QDs to realize  efficient Bell-state measurements \cite{Borregaard2019} or photon-photon quantum gates \cite{Schrinski2021}. Taking a broader perspective, the ability to deterministically couple multiple quantum emitters opens a new arena of studying non-equilibrium quantum many-body physics of strongly correlated light and matter, which could be used in quantum simulations of strongly correlated condensed matter systems~\cite{Noh2016}.

\textbf{Acknowledgments:}
The authors thank Xiao-Liu Chu for fruitful discussions at the beginning of the project.  

\textbf{Funding:} 
We gratefully acknowledge financial support from Danmarks Grundforskningsfond (DNRF 139, Hy-Q Center for Hybrid Quantum Networks). 
B.S. acknowledges financial support from Deutsche Forschungsgemeinschaft (DFG, German Research Foundation), Grant No. 449674892. 
O.S. acknowledges funding from the European Union’s Horizon 2020 research and innovation programme under the Marie Skłodowska-Curie grant agreement No. 801199.
S.S., A.L. and A.D.W. acknowledge financial support of the German-French University DFH/UFA within the CDFA-05-06 as well as BMBF QR.X project 16 KISQ 009.

\textbf{Author contributions:} A.T., V.A., and C.J.v.D. carried out the measurements. Y.W. and L.M. designed and fabricated the sample. O.S., B.S. and A.S developed the theoretical model, A.T., V.A and C.J.v.D. analyzed the data and prepared the figures. S.S., A.D.W., and A.L. carried out the growth and design of the wafer. A.T., V.A., C.J.v.D., and P.L. wrote the manuscript with input from all the authors. P.L. supervised the project. 

\textbf{Competing interests:} P.L. is founder of the company Sparrow Quantum that commercializes single-photon sources.

\textbf{Data and materials availability:} The data presented in the main text and Supplementary Materials  for this publication are openly available from \cite{SMerda}.

\bibliographystyle{apsrev4-1} % this is necessary for multibib to work with document class article, otherwise the references won't show up!
\let\oldaddcontentsline\addcontentsline% Store \addcontentsline
\renewcommand{\addcontentsline}[3]{}% Make \addcontentsline a no-op
\bibliography{arxiv_2QD.bbl}
\let\addcontentsline\oldaddcontentsline% Restore \addcontentsline

%%%%%%%%% Supplementary materials %%%%%%%%%%
\pagebreak
\clearpage 

\title{Supplementary materials for \\ \TitleName}

\maketitle
\onecolumngrid

\setcounter{equation}{0}
\setcounter{figure}{0}
\setcounter{table}{0}
\setcounter{page}{1}
\makeatletter
\renewcommand{\theequation}{S\arabic{equation}}
\renewcommand{\thefigure}{S\arabic{figure}}
\renewcommand{\thetable}{S\arabic{table}}
\renewcommand{\bibnumfmt}[1]{[S#1]}
\renewcommand{\@seccntformat}[1]{%
  \csname the#1\endcsname.\quad
}

{
  \hypersetup{linkcolor=blue}
  \tableofcontents
}

\setcounter{secnumdepth}{3}

\section{Theoretical model}
\label{sec:SMtheory}
The theoretical description is based on the master equation for the system, which reads $(\hbar =1)$
\begin{align}
\label{eq:rhodot}
  \dot \rho = \mathcal{L}_\mathrm{tot}[\rho]= -i \left[H , \rho \right]
  %-i \left[\mathcal{H}_\text{eff} \rho - \rho \mathcal{H}_\text{eff}^{\dag}\right] 
+   \mathcal{L}_\text{coup}[\rho]
  + \mathcal{L}_\text{decay}[\rho] + \mathcal{L}_\text{deph}[\rho].
\end{align}
Here the Hamiltonian \(H =\sum_m \Delta_m\sigma^z_m/2+ \Omega_m(e^{i\theta_m}\sigma_m^{+}+e^{-i\theta_m}\sigma_m^{-})/2\) accounts for the detuning and potential driving of the involved emitters with relative phases $\theta_m-\theta_n$.
In our treatment, we include three different Liouvillians. The first one $\mathcal{L}_\mathrm{coup}[\rho]$ describes the waveguide mediated coupling. This term is the main focus of this work and  is discussed in detail below. Additionally, we include  a Liouvillian  corresponding to the decay of the emitters to modes other than the waveguide (side modes) 
\begin{align}
  \mathcal{L}_\text{decay}[\rho] = 
   %\underbrace{\sum_{m,n} \Gamma_{m,n} \sigma_{m}^{-} \rho \sigma_{n}^{+}}_\text{waveguide coupling} 
  %+ 
 % \underbrace{
 \sum_{m} \gamma^{s}_m \sigma_{m}^{-} \rho \sigma_{m}^{+}
 %}_\text{non-waveguide coupling (side)}
\end{align}
with a decay rate $\gamma^{s}_m$ for the $m$th emitter. 
Finally we also include dephasing  described by
\begin{align}
\mathcal{L}_\text{deph}[\rho]&= \frac{\gamma_d}{2}\sum_{m}\left[ \sigma^z_{m} \rho \sigma_{m}^z - \rho\right], 
\end{align}
where the dephasing rate $\gamma_d$ is assumed similar for both emitters for simplicity.
This is the standard description of pure dephasing  acting individually on each emitter. 

To describe the waveguide mediated coupling, we use the well-known rotating-wave approximated Jaynes-Cummings coupling Hamiltonian in one dimension $H'=\sum_m\int \mathrm{d}k (g_{k,m}\sigma_m^+a_k+g^*_{k,m}a_k^\dagger\sigma_m^-)$ with light mode creation operators $a^\dagger$ and Pauli operators $\sigma_m^+=|e\rangle_m\langle g|_m$ for different QD sites labeled with $m=1,2,3$. The coupling of the light modes to the dipoles is given by $g_{k,m}=-\mathbf{E}_k(\textbf{r}_m)\cdot \mathbf{d}_m$, i.e., the electrical field mode solution in the waveguide at coordinate $\mathbf{r}_m$ times the dipole vector of the $m$th quantum dot (QD). 

The experiment is conducted in mirror-symmetric photonic-crystal waveguides (PCWs) that have primarily linear local polarization so that effects of directional chiral light-matter coupling can be ignored \cite{Lodahl2017}, i.e., coupling coefficients for left and right propagation are of equal magnitude: $|g_{k,m}|\approx |g_{-k,m}|$. The absence of chirality assumed here is experimentally confirmed by comparing amplitudes of resonant transmission dips (\figref{fig:SMbfield}) while sweeping the external magnetic field in two opposite directions. For this, we look at the amplitude ratios between two dipoles for each QD as a function of external magnetic field (\figref{fig:SMchiral}). We repeated the same measurement with the direction of the magnetic field reversed (\figref{fig:SMchiral}). The results show `mirror' symmetry around zero magnetic field up to 4 T, confirming the assumption of non-chiral interaction.

According to Bloch's theorem, we can write the electric field in the periodic waveguide as $\mathbf{E}_k(\mathbf{r})=\mathbf{U}_k(\mathbf{r})e^{ikz}$ with a periodic function  $\mathbf{U}_k(\mathbf{r})$. As we argue in Sec. \ref{sec:offres} below, the procedure used to select QDs favors QDs located at similar positions within the unit cell so that $\mathbf{U}_k(\mathbf{r}_m)$ is similar  for all QDs. In combination with the linear polarization, this means that apart from an overall phase, which can be absorbed by changing the origin of the $z$-axis, we can choose the mode function to be real $\mathbf{U}_k(\mathbf{r}_m)\in \mathbb{R}^3$ for all QDs. 
Since the waveguide obeys time-reversal symmetry, the mode functions must fulfill $\mathbf{E}_k=\mathbf{E}_{-k}^*$. This means that the periodic function $\mathbf{U}_k(\mathbf{r}_m)$ is also real for backward propagation since  $\mathbf{E}_k=\mathbf{E}_{-k}^*$ implies $\mathbf{U}_{-k}=\mathbf{U}_k$.
Furthermore, apart from the propagation phases ($e^{ikz}$), the fact that $\mathbf{U}_k(\mathbf{r}_m)$ is real means that any phase of the coupling constant can only arise from the dipole moment of the transition. Since this dipole moment is the same for the coupling to all modes, any phase in the dipole moment can be absorbed by a redefinition of $\sigma_m^-$.  We thus choose real coupling constants in the following, and the phases that appear will be due to propagation. It is noted that any net phases due to the QDs not having the same position within the unit cell can be incorporated into the definition of the phase $\phi_{mn}$ below, but the assumption of real coupling constants simplifies the description. 

Integrating out the photonic degrees of freedom in the Markov approximation leads to the Liouvillian for the waveguide mediated coupling \cite{dung2002resonant,asenjo2017exponential,Albrecht2019}
\begin{align}
\mathcal{L}_\mathrm{coup}[\rho] = i\sum_{mn}J_{mn}[\sigma_n^+\sigma_m^-,\rho]-\sum_{mn}\Gamma_{mn}\left[\sigma_m^-\rho\sigma_n^+-\frac{1}{2}\{\sigma_n^+\sigma_m^-,\rho\}\right]    
\label{eq:Lcoup}
\end{align}
with the couplings 
\begin{align}
\label{eq:GJ}
\Gamma_{mn}=\frac{4\pi |g_{k_0,m}||g_{k_0,n}|}{v_g}\, \mathrm{Re}\{e^{i\phi_{mn}}\}%e^{\mathrm{sgn(n-m)}i\theta_{mn}}
\quad\mathrm{and}\quad
J_{mn}=\frac{2\pi |g_{k_0,m}||g_{k_0,n}|}{v_g}\, \mathrm{Im}\{e^{i\phi_{mn}}\}%e^{\mathrm{sgn(n-m)}i\theta_{mn}}
,
\end{align}
which corresponds to the expressions in the main text with the identification $\Gamma_m\beta_m=4\pi|g_{k_0,m}|^2/v_g$, where $v_g$ is the group velocity in the waveguide.
Here, the coupling depends on quantities evaluated at the resonant wavenumber $k_0$. To ease the notation, we omit the subscript and simply denote it $k$ in the main text and below. The nature of the interaction between the emitters depends crucially on the propagation phase $ \phi_{mn}=k|x_{mn}|$ with $|x_{mn}|$ being the distance between emitters $m$ and $n$.  
This phase describes the interference of excitations switching  back and forth  between the emitters and determines whether the interaction is dispersive $|\Gamma_{mn}|\ll |J_{mn}|$ for $\phi_{mn}\approx (N+1/2)\pi$ or dissipative $|J_{mn}|\ll |\Gamma_{mn}|$ for $\phi_{mn}\approx N\pi$ with $N$ being an integer. 

In addition to the master equation for the QDs, we also need to relate the density matrix to the outgoing fields. These  can be determined  by the well-known input-output formalism \cite{Chang2007}.  In case of excitation from the side, i.e., not through the waveguide, the left- and right-going fields are given by (assuming the three QDs are labeled in ascending order from left to right)
\begin{align}
\label{eq:4}
E_L = \sum_n e^{i\phi_{1n}} E_n
\quad\mathrm{and}\quad
E_R = \sum_n e^{i\phi_{n3}} E_n, 
\end{align} 
where $E_n =- i \frac{\sqrt{\beta_n\Gamma_{n}}\sigma_n^{-}}{\sqrt{2}}$.
The intensities are given by the squared modulus of the fields  $ I_{L/R} = |E_{L/R}|^{2}$. 

The theoretical results depicted throughout the main text are obtained via  numerical integration of Eq.~\eqref{eq:rhodot}. To account for  spectral diffusion,  we average all results over a normal distribution of detunings $\Delta_m$ with standard deviation $\sigma_{sd}$. Alternatively, for a simplified analytical treatment, we can ignore dephasing and only consider the dynamics in the single excitation subspace. In this case, the master equation \eqref{eq:rhodot} reduces to
\begin{align}
\label{eq:rhodot_simple}
  \dot \rho = 
  -i \left[\mathcal{H}_\text{eff} \rho - \rho \mathcal{H}_\text{eff}^{\dag}\right] 
\end{align}
with  the effective Hamiltonian given in Eq. \eqref{eq:EffHamiltonian} of the main text. This evolution corresponds to the so-called no-jump evolution of the quantum Monte-Carlo wavefunction approach. The omitted jump terms, representing the state after a decay,  prepare the system in the joint ground state, which doesn't emit any light. We can therefore understand the full system dynamics and the emitted light from the no-jump evolution. On the other hand, when the system starts in the doubly excited state, it can be reexcited or undergoes dephasing, the jump evolution is important, and we resort to the full master equation  \eqref{eq:rhodot} for a complete description of the dynamics.

To understand the properties of the emitted light, we now turn to  general symmetry properties that arise for pure dissipative coupling, i.e.\, $\phi_{mn}=N\pi$; the experimental scenario  observed here.
We write the intensities as
\begin{align}
& I_L = \sum_{mn} E^*_m E_n e^{-i \phi_{mn}}, \\
& I_R = \sum_{mn}  E^*_m E_n e^{i \phi_{mn}}.
\end{align}
For purely dissipative coupling we thus have $I_L = I_R$ since for $\phi_{mn}=N\pi$ the exponential terms $e^{i \phi_{mn}} =e^{-i \phi_{mn}} = (-1)^N$, making the signal equal at both detectors.

The symmetry ($\Delta_{mn}\rightarrow -\Delta_{mn}$) in detuning $\Delta_{mn}=\Delta_m-\Delta_n$ between the different QDs can be explained by observing that in case of purely dissipative coupling, the only imaginary contribution to the master equation \eqref{eq:rhodot}, apart from potential phases in the driving, arises from terms $\propto\Delta_m\sigma_m^z$. Thus, changing $\Delta_m \to -\Delta_m$ leads to complex conjugation of the Liouvillian $\mathcal{L}_\mathrm{tot}\to\mathcal{L}_\mathrm{tot}^*$, and we have $e^{\mathcal{L}_\mathrm{tot}^*t}[\rho^*]=(e^{\mathcal{L}_\mathrm{tot}t}[\rho])^*$, i.e., the master equation is solved by the complex conjugated density matrix $\rho^*$. Since $\rho(t)\to\rho^*(t)$ does not affect measurement results  if Eq. \eqref{eq:4} is real ($\phi_{mn}=N\pi$), it follows that for initially real $\rho=\rho^*$, e.g., for a QD quickly excited to the excited state, the time evolution will produce the same intensity for opposite detunings. On the other hand,  driving two QDs simultaneously can induce a complex phase between the QDs  that breaks the symmetry, as is experimentally observed and discussed in the main text.   

\begin{figure}[h!]
\begin{center}
	\includegraphics[width=0.7\linewidth, trim=0.5cm 0.45cm 0.25cm 0.3cm,clip]{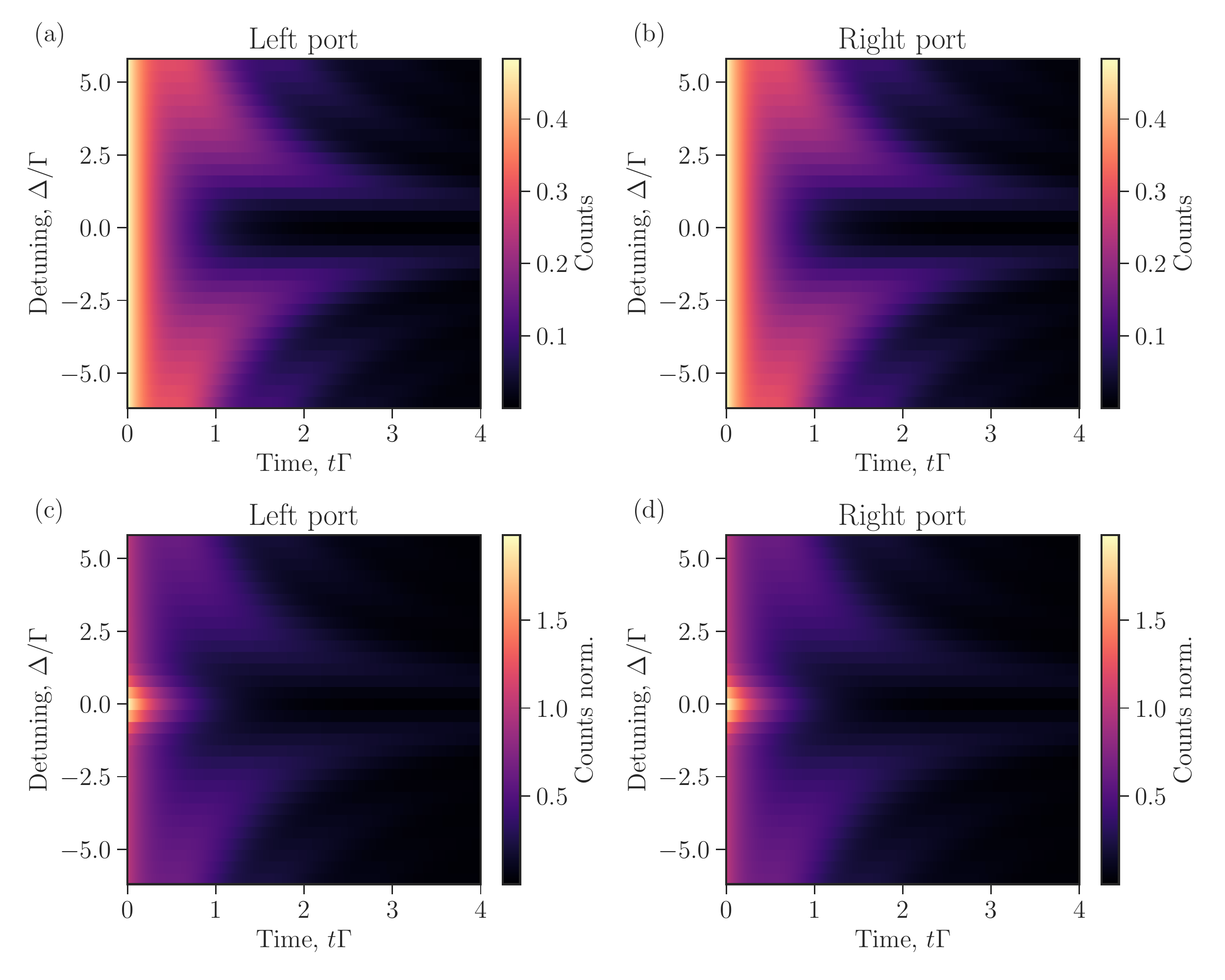}
	\end{center}
	\caption{Exemplary calculated time resolved intensities for left (a,c) and right (b,d) ports according to the analytical results (Eq.~\eqref{eq:OutputfieldsRight}), in qualitative agreement with the experimental data from the main text. The normalization to the sum of each time trace was used for (b,d). The following parameters are assumed: $\phi_{12} = 0$, $\beta_1 = \beta_2 = 0.99$, $\Gamma_1 = \Gamma_2 = \Gamma$, $\gamma_d = 0$. 
}
	\label{fig:SM_ItheoryG}
\end{figure}
\begin{figure}[h!]
\begin{center}
	\includegraphics[width=0.65\linewidth, trim=0.55cm 0.45cm 0.25cm 0.1cm,clip]{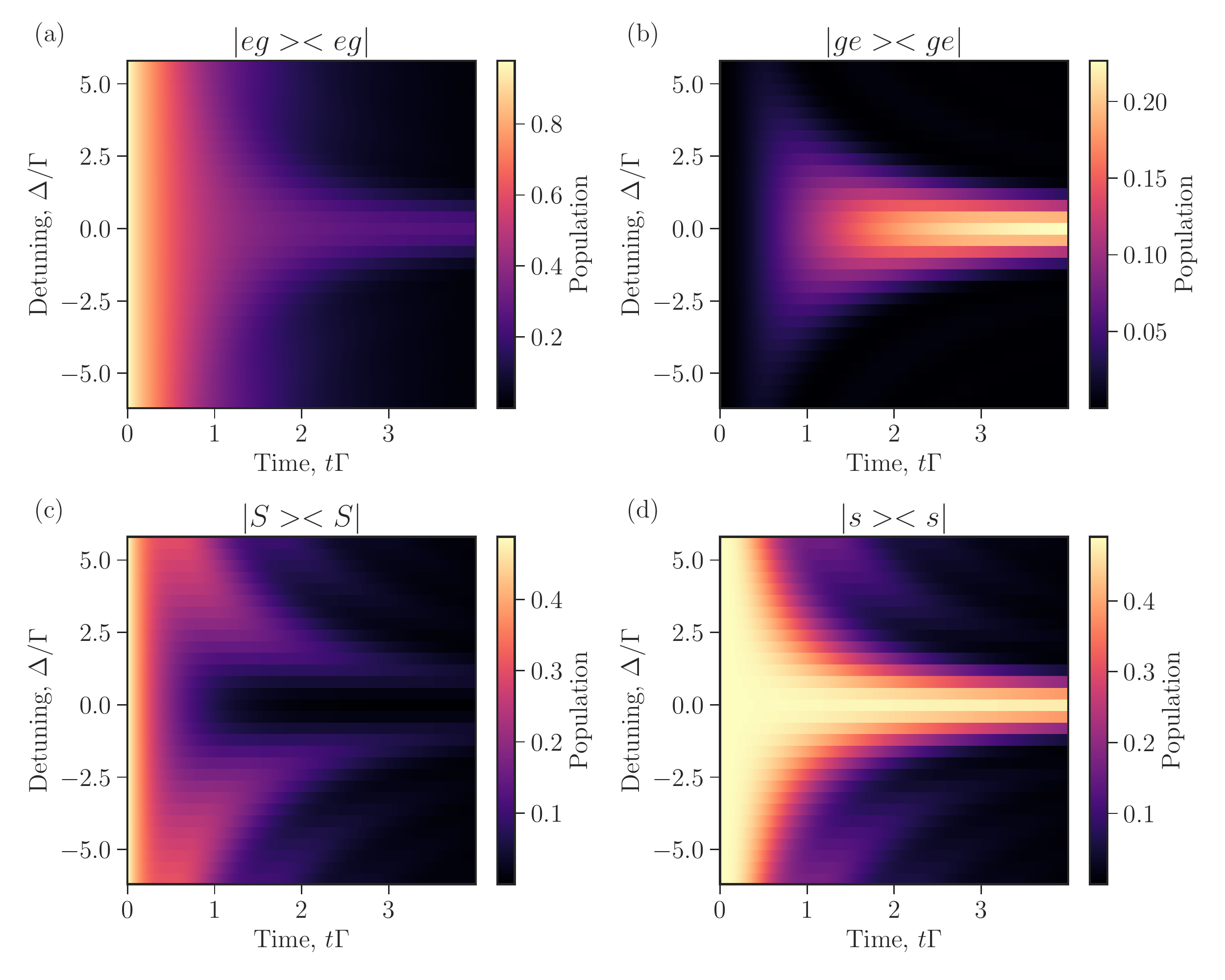}
	\end{center}
	\caption{ 
 Theoretical population of the different density matrix entries (which are not directly accessible in the experiment): Due to the decay of the superradiant state $|S\rangle$, the excitation $|eg\rangle\langle eg|$ gets redistributed to the other QD $|ge\rangle\langle ge|$ in the shape of the subradiant state $|s\rangle$, which exhibits an extremely long lifetime if close to $\Delta=0$. For larger detunings, we observe oscillations between $|S\rangle$ and $|s\rangle$ that quickly die out due to the fast emission of the superradiant state. Here, the same parameters as for \figref{fig:SM_ItheoryG} were used.
}
	\label{fig:SM_NtheoryG}
\end{figure}
\begin{figure}[h!]
\begin{center}
	\includegraphics[width=0.65\linewidth, trim=0.55cm 0.45cm 0.25cm 0.3cm,clip]{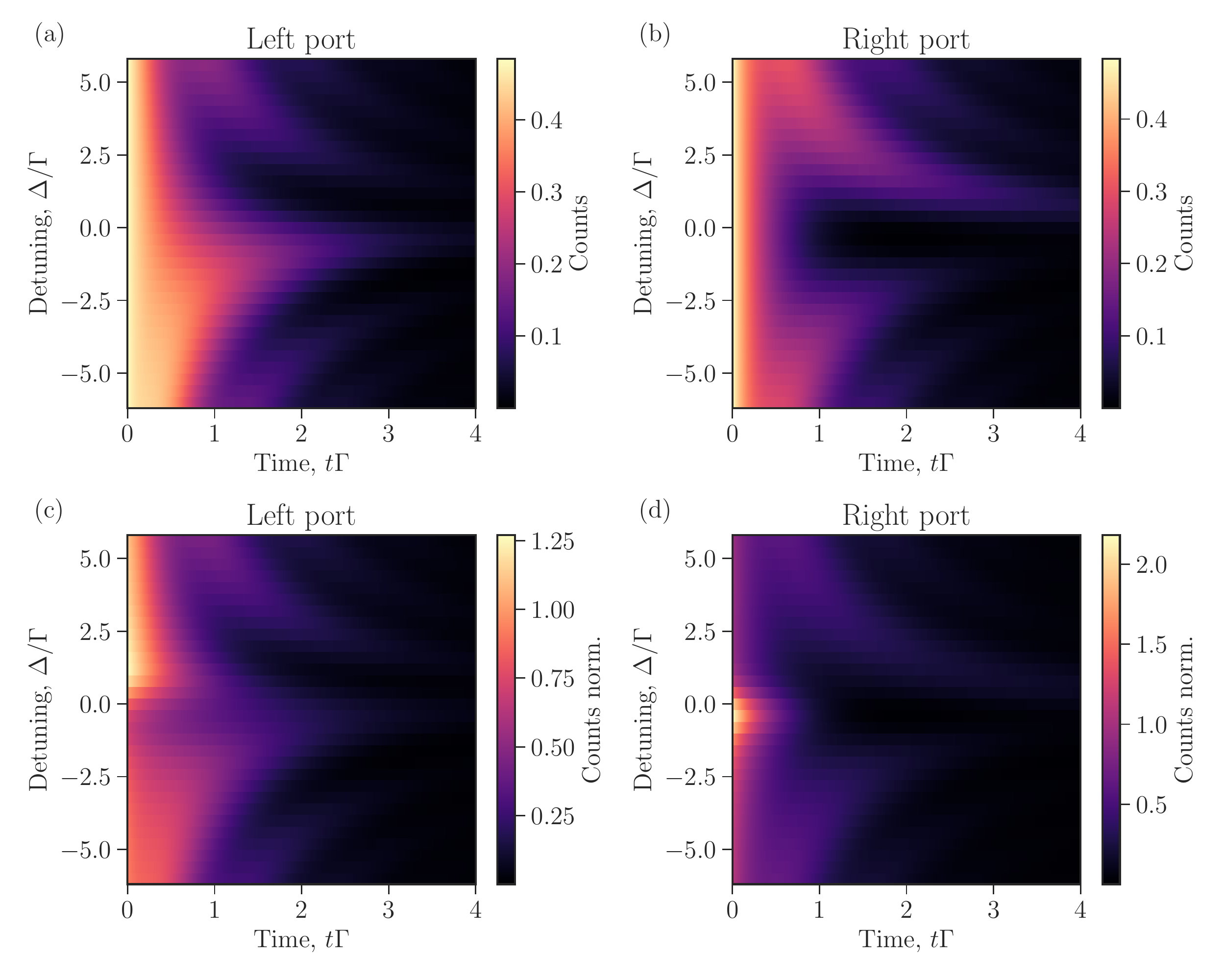}
	\end{center}
	\caption{ Increasing the relative phase $\phi_{12} = \pi/4$ acquired between the two QDs leads to asymmetries within and between the left (a,c) and right (b,d) output ports. The absence of asymmetry between positive and negative detunings in the experimental data is a clearly visible indicator for $\phi_{12} \simeq 0$. The normalization to the sum of each time trace was used for (b,d). The following parameters are assumed: $\phi_{12} = \pi/4$, $\beta_1 = \beta_2 = 0.99$, $\Gamma_1 = \Gamma_2 = \Gamma$, $\gamma_d = 0$.
}
	\label{fig:SM_theoryJ}
\end{figure}

To give analytical expressions for the output fields of two QDs close to resonance, we assume the initial $\pi$-flip (exciting the left QD) to be quasi-instantaneous compared to the lifetimes $1/\Gamma_1,1/\Gamma_2$ and neglect dephasing ($\gamma_d=0$). For this simplified scenario, we can give concise output fields arriving at the right detector at time $t$ 
$E^R(t)=E^R_\mathrm{sup}(t)+E^R_\mathrm{sub}(t)$
with the (original) super- and subradiant contributions
\begin{align}\label{eq:OutputfieldsRight}
E^R_\mathrm{sup}(t)=&\frac{-i\sqrt{\Gamma_1\beta_1}(\Gamma_2\beta_2-i\Delta+(S+\Gamma_1-\Gamma_2)/2)}{\sqrt{2}S}\exp[-(\Gamma_1+\Gamma_2+S)\frac{t}{4}]\nonumber\\
E^R_\mathrm{sub}(t)=&\frac{i\sqrt{\Gamma_1\beta_1}(\Gamma_2\beta_2-i\Delta+(-S+\Gamma_1-\Gamma_2)/2)}{\sqrt{2}S}\exp[-(\Gamma_1+\Gamma_2-S)\frac{t}{4}],
\end{align}
where $S=\sqrt{4e^{2i\phi}\Gamma_1\Gamma_2\beta_1\beta_2+(2i\Delta-\Gamma_1+\Gamma_2)^2}$. We observe that for the ideal setup ($\Gamma_1=\Gamma_2=\Gamma$, $\beta_1=\beta_2=1$, $\Delta=0$, $\phi=2N\pi$) we get $E^R_\mathrm{sub}(t)=0$. This corresponds to the subradiant part of the state $(|e_1,g_2\rangle-|g_1,e_2\rangle)/\sqrt{2}$ being reflected between the two QDs and never leaving the system, while the superradiant part $(|e_1,g_2\rangle+|g_1,e_2\rangle)/\sqrt{2}$ decays with double the decay rate $\Gamma^R_\mathrm{sup}=2\Gamma$. For $\phi=(2N+1)\pi$, we achieve the identical scenario, just the super- and subradiant states and fields change place. The output fields in the left port can be separated into sub- and superradiant contributions as well:
\begin{align}\label{eq:OutputfieldsLeft}
E^L_\mathrm{sup}(t)=&\frac{-i\sqrt{\Gamma_1\beta_1}(e^{2i\phi}\Gamma_2\beta_2-i\Delta+(S+\Gamma_1-\Gamma_2)/2)}{\sqrt{2}S}\exp[-(\Gamma_1+\Gamma_2+S)\frac{t}{4}]\nonumber\\
E^L_\mathrm{sub}(t)=&\frac{i\sqrt{\Gamma_1\beta_1}(e^{2i\phi}\Gamma_2\beta_2-i\Delta+(-S+\Gamma_1-\Gamma_2)/2)}{\sqrt{2}S}\exp[-(\Gamma_1+\Gamma_2-S)\frac{t}{4}],
\end{align}
where the prefactors indicate asymmetry in intensity between the right and left output fields for $\phi\neq 2 N\pi$. 

The decay rate of the subradiant state becomes finite as soon as only one of the parameters $\Delta,\phi$ deviates from the ideal values above. In the case of $\beta<1$ for the symmetric waveguide, the subradiant state also decays, but to non-guided "side modes" that do not contribute to the output field so that $E^R_\mathrm{sub}(t)$ is still zero. To the lowest order in those parameters, the decay rate scales quadratically (assuming for simplicity $\Gamma_1=\Gamma_2=\Gamma$ and $\beta_1=\beta_2=\beta$),
\begin{align}
\Gamma^R_\mathrm{sub}=\Gamma(1-\beta)+\Gamma\beta\phi^2/2+(2-\phi^2)\Delta^2/(4\Gamma\beta)+\mathcal{O}(\Delta^3/\Gamma^3)+\mathcal{O}(\phi^3),
\end{align}
while the rate $\Gamma^R_\mathrm{sup}$ is lowered by the same correction in $\phi$ and $\Delta$.
This manifests in a broad corridor of almost only superradiant decay around $\Delta=0$; see the exemplary theoretical study depicted in \figref{fig:SM_ItheoryG} qualitatively agreeing with the actual experimental results shown in the main text. The oscillations in intensity for increasing $\Delta$ stem from the oscillations between the sub- and superradiant states: in a simplified picture, a phase is acquired $|e_1,g_2\rangle+e^{i\Delta t}|g_1,e_2\rangle$, letting both states evolve around the Bloch sphere (\figref{fig:3}) and turn into one another. This is depicted in \figref{fig:SM_NtheoryG} for the same theoretically ideal scenario showing an oscillation between the super- and subradiant states $|S\rangle,|s\rangle$. A finite driving time, much shorter than the emitter lifetimes $\Gamma_i$, is assumed during this calculation. The almost perfect symmetry of the decay tails also indicates a relative phase $\phi\simeq0$ since this oscillation is corrected in lowest order as $(\Delta+\Gamma\beta\phi)/2$, increasing or decreasing the oscillation frequency depending on the signs of $\Delta$ and $\phi$, see e.g., \figref{fig:SM_theoryJ}.

\section{Sample and optical setup}
\label{sec:SMsetup} 
\begin{figure}[h!]
\begin{center}
	\includegraphics[width=0.8\linewidth]{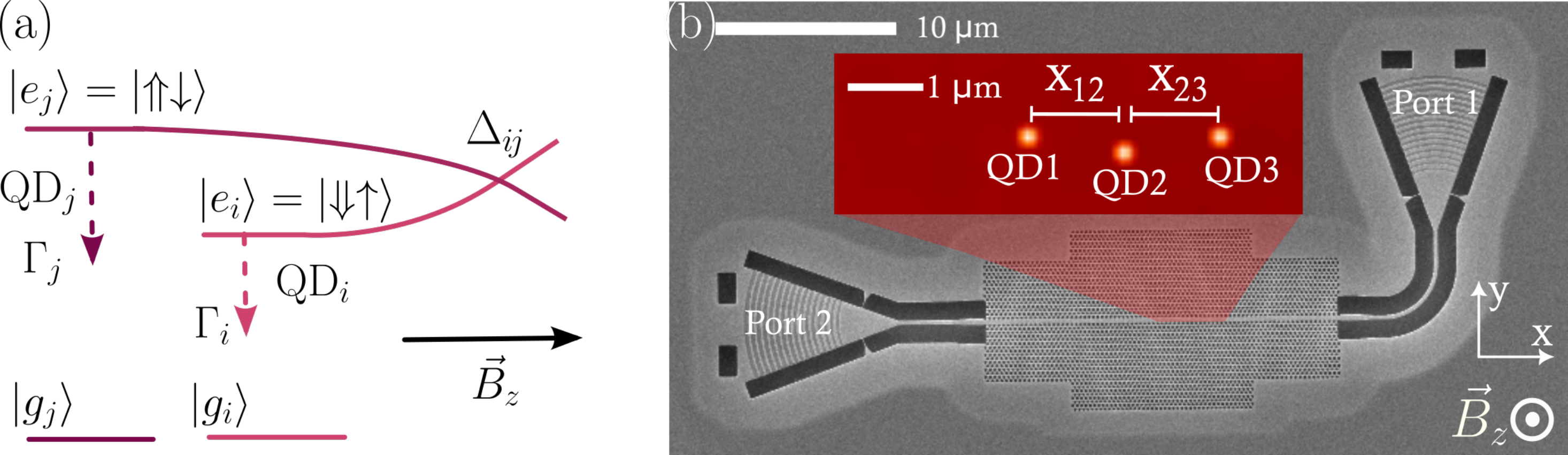}
	\end{center}
	\caption{ 
	(color online) 
	(a) Energy level diagram of two QDs detuned by $\Delta_{ij}$ and controlled by an external magnetic field $B_z$.
	(b) Scanning electron microscope image of the equivalent two-sided PCW structure. The QD is addressed by lasers propagating from free space, and the emitted photons are coupled into the waveguide mode, scattered out from right (port 1) or left (port 2) grating couplers, and coupled into a single-mode fiber. The inset contains the fluorescence image of the QDs coupled to the PCW (cf. Section \ref{sec:SMimag} below).
}
	\label{fig:SMSEM}
\end{figure}

The sample similar to the one described in \cite{Uppu2020} is cooled to 4 K in a closed-cycle cryostat with optical access along the $z$-direction, see  \figref{fig:SMSEM}. A vector magnet allows us to apply an external magnetic field up to 5~T in the $z$ direction and 3 T in the $x$-direction. 
A confocal microscope with 0.8 numerical aperture was used to image the sample, excite the QD from free space, and couple light into the waveguide.  Polarization control is used to reject the laser scattering while collecting the $x$ or $y$-polarized emission of collection port 1 and port 2, respectively.  Shallow etched gratings \cite{Zhou2018} couple to linearly polarized near-gaussian modes and have above 50 nm (FWHM) bandwidth providing a near-equal broadband collection and excitation of the QDs coupled to the waveguide. Details about the sample growth and fabrication can be found in \cite{Uppu2020}.

Acousto-optic modulators (AOMs) are used to stabilize the optical power of two lasers. 
The continuous-wave laser used for the excitation of QD optical transitions is continuously tunable between 900-980 nm. In addition, a pulsed 5 ps laser is utilized for the excited state lifetime measurements and p-shell excitation. The lasers are combined using non-polarizing beamsplitters (BS) to address the photonic chip inside the cryostat at  4 K temperature. The polarization for each laser is controlled using polarizing beamsplitters (PBS) and a pair of half- ($\lambda$/2) and quarter-waveplates ($\lambda$/4). The emission from the outcoupling grating of the chip is collected using a BS and is sent through a spectral filtering setup. The filtering is done using a reflective diffraction grating with the resulting bandwidth of 25 GHz FWHM, while the resonant excitation is filtered with a transmission diffraction grating and an etalon filter with a bandwidth of approximately 3 GHz. Avalanche single-photon detectors (APD) or superconducting single photon detectors (SNSPD) are used to detect the filtered signal depending on the experiment.

In our experiment, the InAs QDs are embedded into a PCW  made from a suspended GaAs membrane and connected to grating couplers at each of its ends to couple light in and out of the PCW (\figref{fig:SMSEM}).  The intrinsic layer containing self-assembled InAs QDs is grown inside the membrane comprising a p-i-n diode. The charge stabilization allows achieving near lifetime-limited emission linewidths, while the integration into the PCW leads to near-unity $\beta$ factor strongly enhancing the photon-emitter interaction \cite{Arcari2014}. This makes it possible to coherently couple emitters separated by a distance much longer than the wavelength.

\section{Optical spectroscopy}
\label{sec:SMspec}

\begin{figure}[h!]
\begin{center}
	\includegraphics[width=0.6\linewidth]{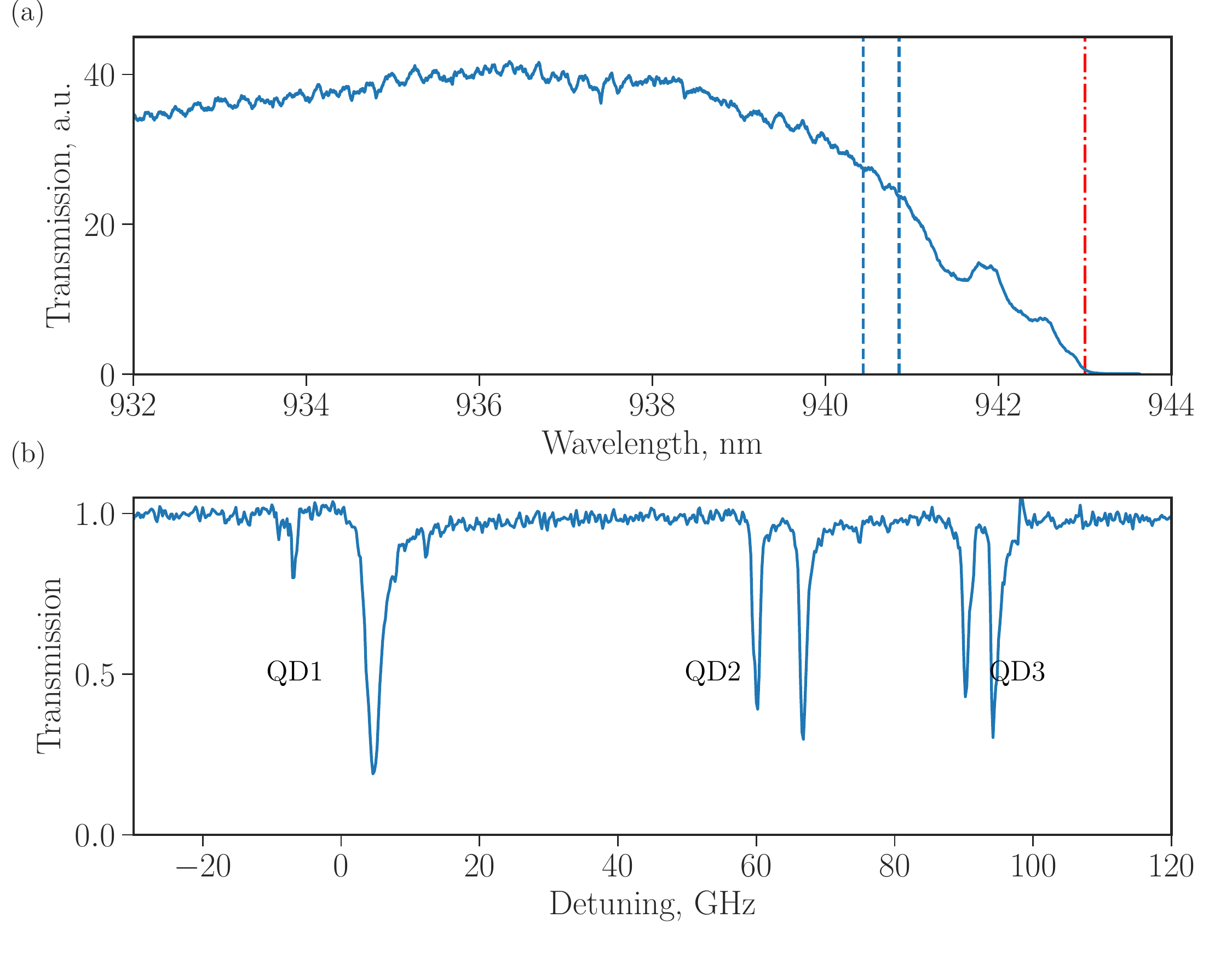}
	\end{center}
	\caption{ 
	(color online) (a) Transmission spectrum of the PCW.  
	(b) Transmission spectrum of the PCW at a bias voltage of 1.24 V showing transmission dips of multiple QDs coupled to the optical mode of the PCW. The spectral region containing the QDs under study is indicated between dashed lines centered around 940.55~nm. The band edge of the PCW is estimated to be around 943 nm (dash-dotted line).
}
	\label{fig:SMpcw}
\end{figure}

\figref{fig:SMpcw} shows the transmission spectra of the PCW and resonant transmission dips of QDs under study with respect to the PCW band edge. These spectra were taken at zero magnetic field and at 1.24~V bias voltage. 

\begin{figure}[h!]
\begin{center}
	\includegraphics[width=0.7\linewidth]{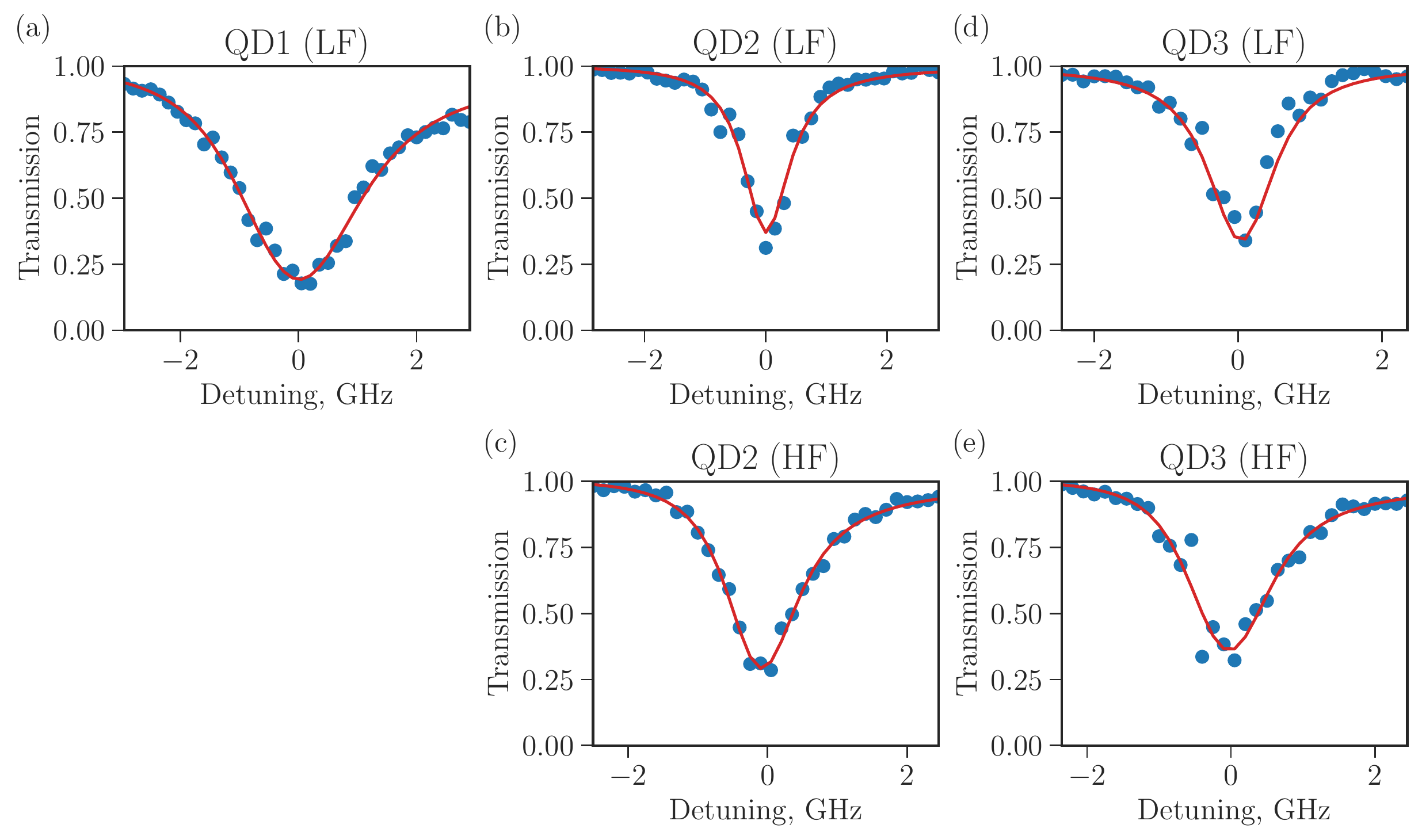}
	\end{center}
	\caption{ 
	(color online) Fitting of the resonant transmission spectra for QD1 (a), QD2 (b,c) and QD3 (d,e) for the high frequency (HF) and low frequency (LF) dipoles for each QD at zero magnetic field $B=0$. For QD1 only the LF dipole is visible at $B=0$.
}
	\label{fig:SMRTfit}
\end{figure} 

\begin{figure}[h!]
\begin{center}
	\includegraphics[width=0.73\linewidth, trim=0.65cm 0.55cm 0.35cm 0.4cm,clip]{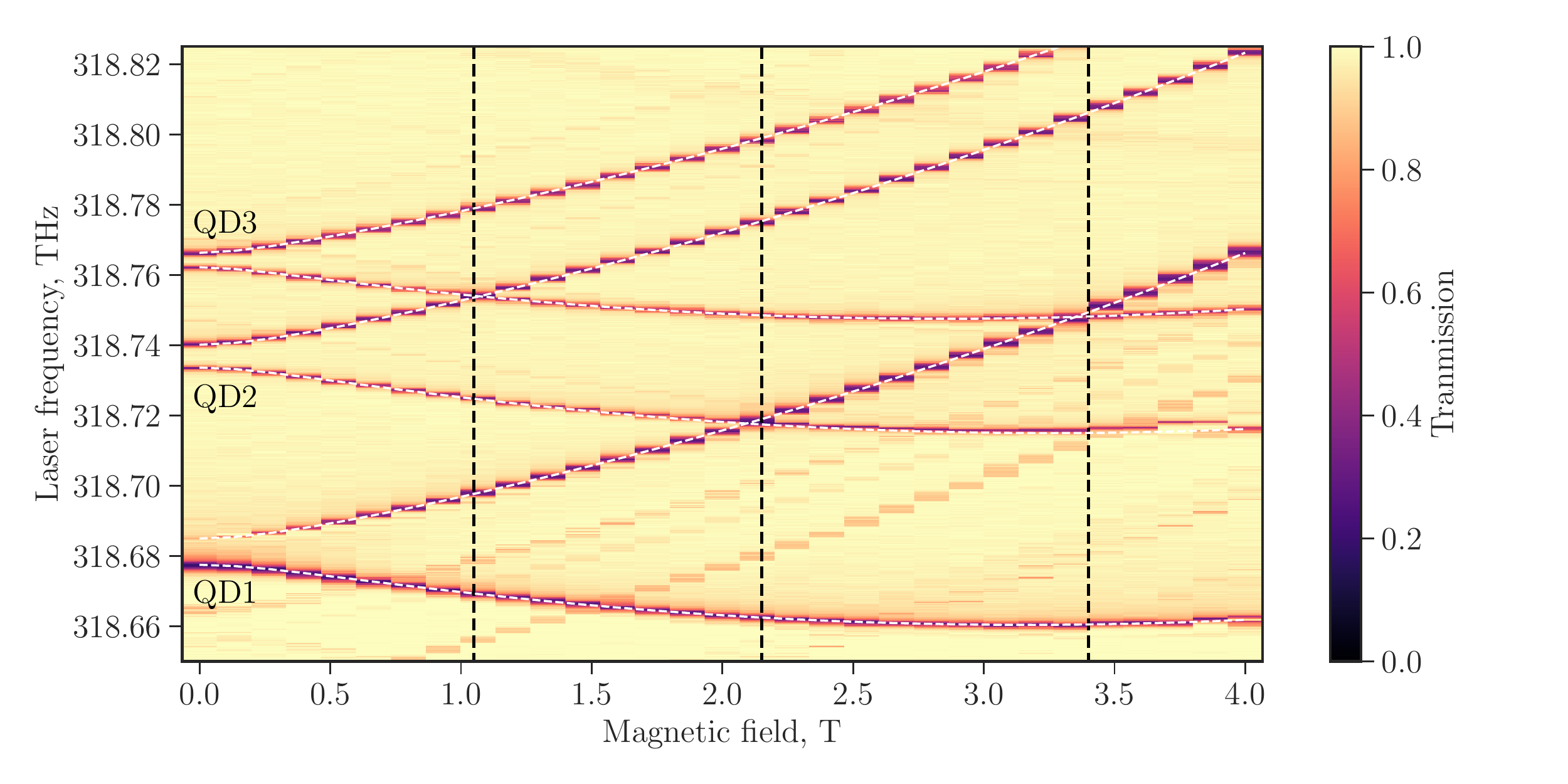}
	\end{center}
	\caption{ 
	(color online)  The transmission spectrum of the PCW as a function of the out-of-plane external magnetic field $B$ (Faraday configuration). The white dashed lines are the theoretical prediction for the Zeeman effect. The diamagnetic shift $\sigma B^2$ is estimated to be $\sigma = 2.05$ GHz T$^{-2}$.
}
	\label{fig:SMbfield}
\end{figure}

In \figref{fig:SMbfield}, the laser frequency and external magnetic field are scanned. At each laser frequency, laser background correction is performed by normalizing to the  counts measured at the nonresonant bias voltage of 1 V.

All QDs at zero magnetic field show fine structure splitting (FSS) to be 7.5~GHz (QD1), 6.5~GHz (QD2), and 4.1~GHz (QD3), respectively. The measured FSS corresponds well to the average value for QDs in bulk InAs.  QD1 shows strong asymmetry for the  two $x$ and $y$ dipoles, where only the high frequency (HF) dipole is well coupled to the PCW at zero magnetic field, while QD2 and QD3 show a similar coupling also for the low frequency (LF) dipole. These observations are due to the spatial variations of the polarization of the PCW mode giving different projections of the QD transition dipole moments~\cite{Lodahl2015}.

The situation changes once the external magnetic field is applied in the z-direction. Two dipoles are split by the Zeeman interaction, and the relative ratio in amplitude between the two dipoles is equalized. This is due to the fact that for an out-of-plane magnetic field, the exciton transition dipoles become circularly polarized, and therefore the projection on the local linear PCW polarization implies that both dipoles couple equally. At the high field, the Zeeman effect is strongly nonlinear because of the diamagnetic shift.

\begin{table}[h]
\caption{ 
Summary of the extracted QD parameters at zero magnetic field:  fine structure splitting (FSS), the exciton gyromagnetic ratio (g-factor) $g_z$, radiative recombination rate $\Gamma$,  spectral diffusion broadening $\sigma_{sd}$.}
\label{tab:QDsum}
\setlength{\tabcolsep}{5pt}
\renewcommand{\arraystretch}{1.4}
\begin{center}
\begin{tabular}{c|c|c|c}
\toprule[1pt]\midrule[0.3pt]
          & QD1 & QD2  & QD3   \\ \hline
FSS, GHz & 7.5  & 6.5  & 4.1  \\ 
$g_z$ & 1.86  & 1.91 & 1.67  \\ 
$\Gamma/2\pi$, GHz & 2.3 (LF)   & 0.87 (HF), 0.69 (LF)  & 0.92 (HF), 0.72 (LF)  \\
$\beta$ & 0.94 (LF)   & 0.88 (HF), 0.83 (LF)  & 0.84 (HF), 0.83 (LF)  \\
$\sigma_{sd}/2\pi$, GHz & 0.59  & 0.22  & 0.6 \\

\midrule[0.3pt]\bottomrule[1pt]  
\end{tabular}
\end{center}
\end{table}

The tuning is done by the out-of-plane (Faraday geometry) magnetic field (\figref{fig:SMbfield}). The external field  gives rise to a pair of Zeeman-split energy levels for the excited state $\ket{\uparrow\Downarrow}$ and $\ket{\downarrow\Uparrow}$ connected to the ``empty" ground state $\ket{g} = \ket{0}$, representing two optical dipoles of the QD. The excited state splitting is used to tune the optical transition in resonance with other QDs. The external magnetic field leads to circular polarizations for both dipoles $\sigma_-$ and $\sigma_+$ for $\ket{0} \longleftrightarrow \ket{\uparrow\Downarrow}$ and $\ket{0} \longleftrightarrow \ket{\downarrow\Uparrow}$, respectively. Using the magnetic field, we are able to tune QD2 and QD3 optical transitions in resonance around $B_z = 1.05$~T, QD1 and QD2 close to $B_z = 2.15$~T and QD1 and QD3 at $B_z = 3.4$~T.

\begin{figure}[h!]
\begin{center}
	\includegraphics[width=0.93\linewidth]{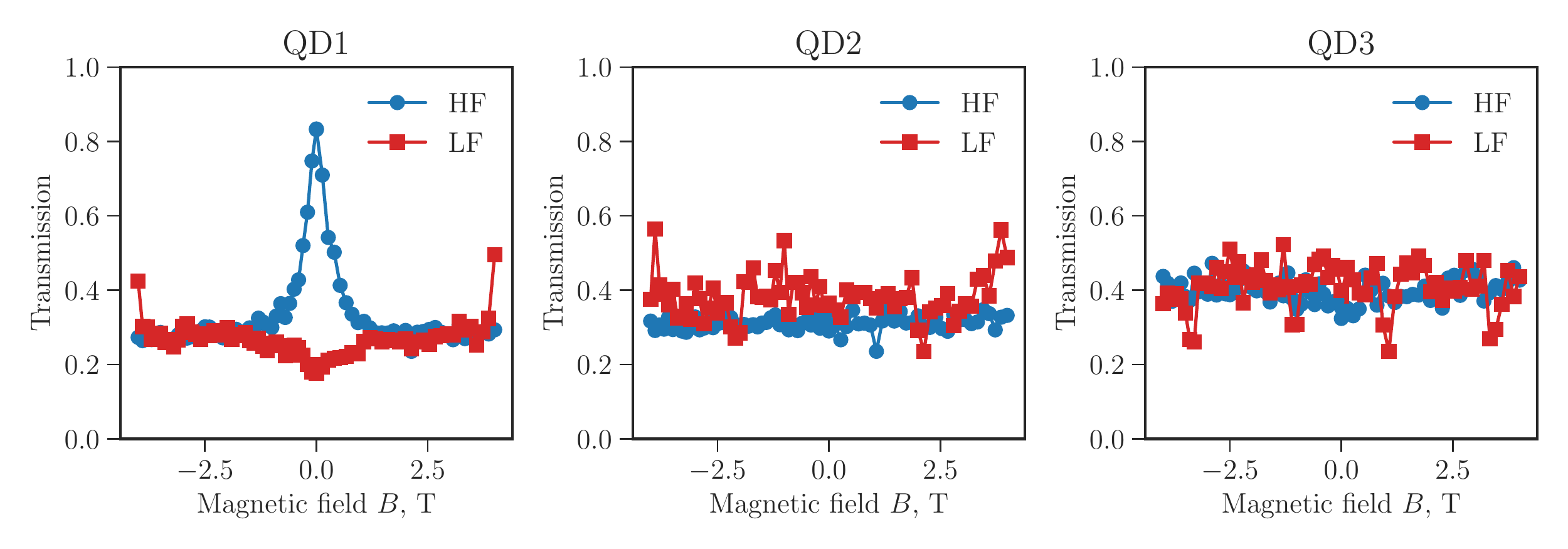}
	\end{center}
	\caption{ 
	(color online)  Minimum of the transmission spectrum as a function of the magnetic field $B$ for high (HF) and  low frequency (LF) dipole for each QD. The symmetry around the zero magnetic field illustrates the non-chiral coupling of each QD to the PCW mode.
}
	\label{fig:SMchiral}
\end{figure}

The absence of chirality is probed experimentally by comparing amplitudes of resonant transmission dips (\figref{fig:SMchiral}) for the external magnetic field sweeps in two opposite directions. The  `mirror' symmetry around zero magnetic field confirms the assumption of non-chiral interaction.

The fitting of the resonant transmission spectra at zero magnetic field using the theoretical model allows extracting the $\beta$-factor to be above $\beta > 0.8$ for all measured dipoles (\tabref{tab:QDsum}). 

\begin{figure}[h!]
\begin{center}
	\includegraphics[width=0.93\linewidth, trim=0.65cm 0.55cm 0.35cm 0.4cm,clip]{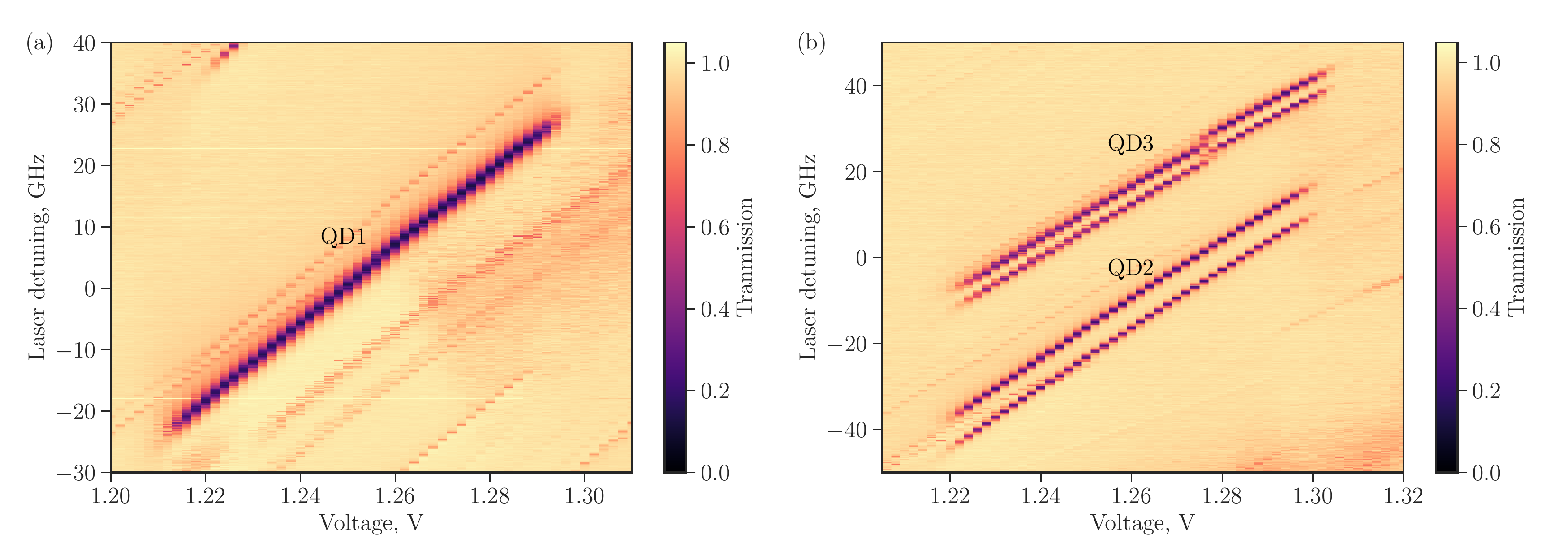}
	\end{center}
	\caption{ 
	(color online) The frequency-voltage scan of the resonance transmission for (a)  QD1, (b) QD2 and QD3  at zero magnetic field with 318.67 THz (a) and 318.75 THz (b) laser frequency offset. 
}
	\label{fig:SMplateau}
\end{figure}

In \figref{fig:SMplateau}, the laser frequency and the bias voltage are scanned to probe the frequency-voltage map of the QD. At each laser frequency, laser background correction is performed to normalize the counts measured at a non-resonant bias voltage of 1~V. All QDs show stable excitons between 1.22~V and 1.3~V. The parallel tuning of the two dipoles for each QD confirms the origin of the measured optical lines. If not stated explicitly otherwise, the  voltage used for all the measurements presented below was 1.24~V.

\begin{table}
\caption{\textbf{Parameters of the photon-mediated coupling for different QD pairs}:  radiative recombination rates $\{\Gamma_{i}, \Gamma_{j}\}$,  spectral diffusion linewidth $\sigma_{sd}$, phonon dephasing rate $\gamma_d$, phase lag $\phi_{ij}$, dissipative coupling rate $\Gamma_{ij}$ and dispersive coupling rate $J_{ij}$.}
\label{tab:1}
\setlength{\tabcolsep}{5pt}
\renewcommand{\arraystretch}{1.4}
\begin{center}
\begin{tabular}{c|c|c|c}
\toprule[1pt]\midrule[0.3pt]
          & QD2-QD3 & QD1-QD2  & QD1-QD3   \\ \hline
$\{\Gamma_{i}, \Gamma_{j}\}/2\pi$, GHz & 0.79, 0.73  & 0.85, 0.8 & 0.9, 0.65  \\
$\sigma_{sd}/2\pi$, GHz & 0.38 & 0.18 & 0.33 \\
$\gamma_d/2\pi$, GHz & 0.03 & 0.03 & 0.03 \\  
$\phi_{ij}$, rad &  0.05      & 0.08 & 0.05\\ 
$\Gamma_{ij}/2\pi$, GHz &  0.61      & 0.66 & 0.61 \\  
$J_{ij}/2\pi$, GHz &  0.03      & 0.05 & 0.03\\  
\midrule[0.3pt]\bottomrule[1pt]  
\end{tabular}
\end{center}
\end{table}

\section{QD imaging}
\label{sec:SMimag}
\begin{figure}[!hb]
\begin{center}
\includegraphics[width=\linewidth, trim=0.65cm 0.55cm 0.35cm 0.4cm,clip]{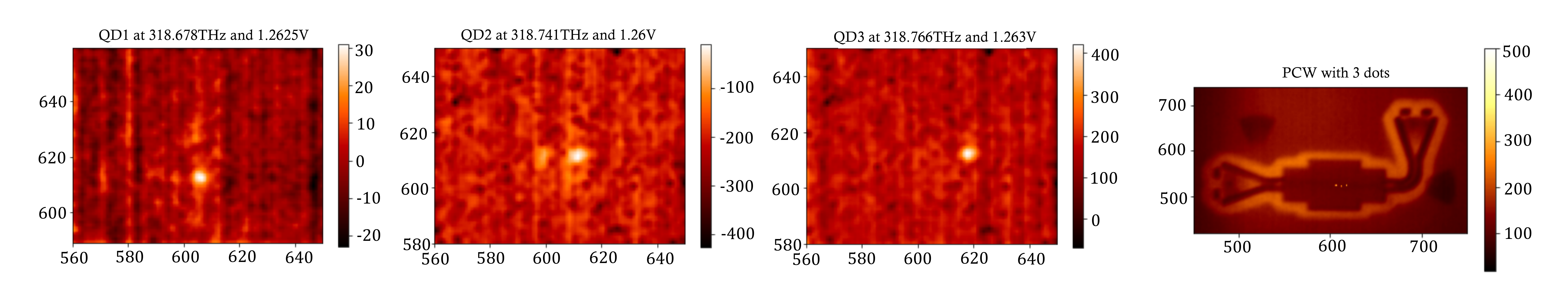}
\end{center}
\caption{Imaging of QD1, QD2, and QD3 in the PCW. The first three (from left) images were taken during resonant transmission measurements and were fitted accordingly to extract the maximum of the counts, revealing each QD position. The fourth image is done using  broadband  light centered around 940 nm illuminating the chip. The image is superimposed with the first  three images. The frequency and bias voltage used for each QD are noted. The axes units are in pixels.  
}
\label{fig:SMimaging}
\end{figure}

To locate the spatial positions of the QDs within the PCW (\figref{fig:SMSEM}) we performed imaging of the sample using a CCD camera. For this, six images were taken, one while exciting each QD on and off resonance, respectively. Another image was acquired while imaging the sample surface using broadband light around 940~nm. 
To excite each QD, the resonant laser light was sent through the PCW (port 1). The CCD camera (DCC CMOS camera (1280 x 1024)) detected the light scattered by the QD outside the PCW mode for an acquisition time of $30$ s. An image of the background (measured at 1~V) was then subtracted, see \figref{fig:SMimaging} (first three images from left).  A gaussian fit to the peak  was used to identify the position of each QD in pixels with respect to the camera resolution. The image of the structure, taken with white light on the sample, was used to  calibrate the dimensions and extract relative distances.

The position (in pixels) for each QD was found from the Gaussian fit. To calibrate the distance, we used the size of the PCW from the white light image (in pixels), and its size in $\mu$m from the mask used to fabricate the chip. We got the ratio between the two units: 181(3) nm/px, leading to the corresponding separations of: 1.25(3) $\mu$m for QD1-QD2 (2nd crossing around 2.15~T), 0.96(3) $\mu$m for QD2-QD3 (1st crossing around 1.05~T) and 2.21(4) $\mu$m for QD1-QD3 (3rd crossing around 3.4~T).

\section{`Off-resonant' excitation}
\label{sec:offres}

\begin{figure}[h!]
\begin{center}
	\includegraphics[width=0.9\linewidth]{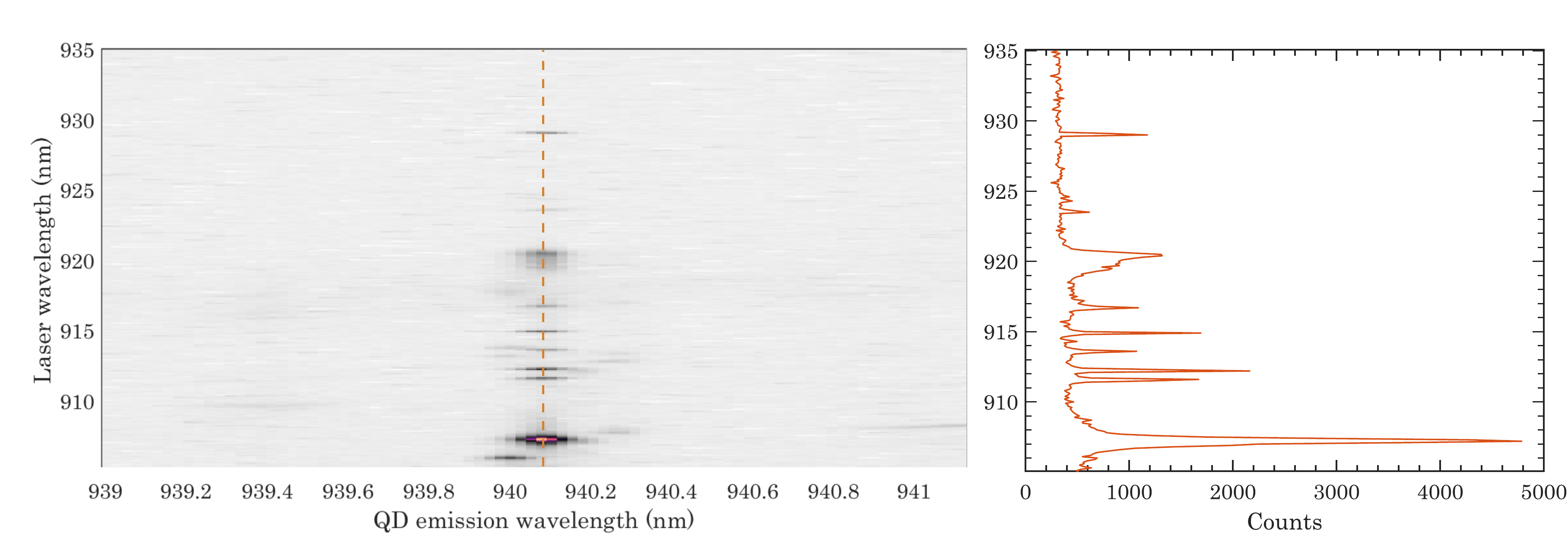}
	\end{center}
	\caption{ Example of the excitation spectra of QD2 at zero magnetic field emitting at 940.09~nm (dashed line). The 2D map on the left side shows higher energy level states of QD2 while the frequency of the continuous  excitation laser is increasing. 
	The excitation around 920.4~nm was used for  lifetime measurements.
  The plot on the right shows photoluminescence excitation spectra recorded at a fixed wavelength of 940.09 nm (dashed line in the left plot).}
	
	\label{fig:SMpshell_spectra}
\end{figure}

\begin{figure}[h!]
\begin{center}
	\includegraphics[width=0.5\linewidth]{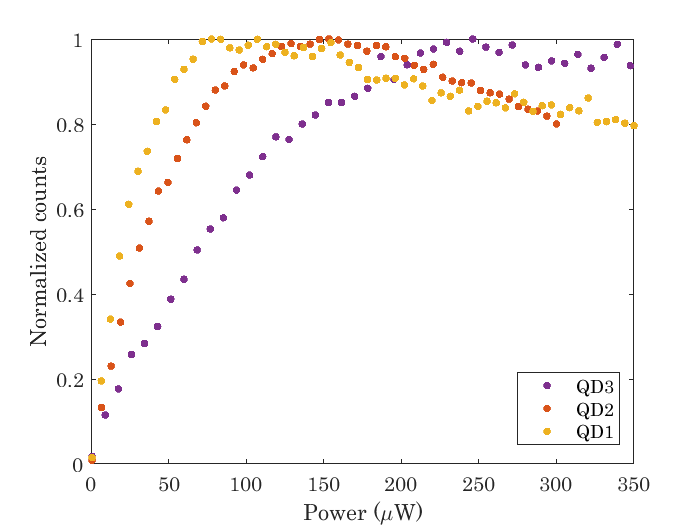}
	\end{center}
	\caption{Power saturation measurements for the three QDs, when driven by a pulsed off-resonant excitation beam. }
	
	\label{fig:SMSaturationpshell}
\end{figure}

To characterize the lifetime of the QDs,  `off-resonant' excitation through higher-order QD shells (p-shell, d-shell~...) was used. For this, a 5 ps 80~MHz pulsed laser was used. The  initial characterization was performed by measuring the excitation spectra of the QD using a continuous laser. An example of the excitation spectra for QD2 is shown on \figref{fig:SMpshell_spectra}. The absence of other QD spectral lines in the recorded spectra guaranteed single QD excitation with this method. The excitation frequency was then chosen such that it was far enough from the resonant one, so that it is easier to filter it out of the collection path, but close enough, so that the relaxation time to the s-shell of the conduction band, would not influence the lifetime measurement itself. Specifically, an excitation around 325.72~THz was used for the lifetime measurement (920.4~nm corresponding to one of the peaks in the figure). The power dependence of all three QDs shows the saturation behavior (\figref{fig:SMSaturationpshell}). An excitation power close to saturation was used during the lifetime measurements.

\begin{figure}[h!]
\begin{center}
	\includegraphics[width=0.9\linewidth]{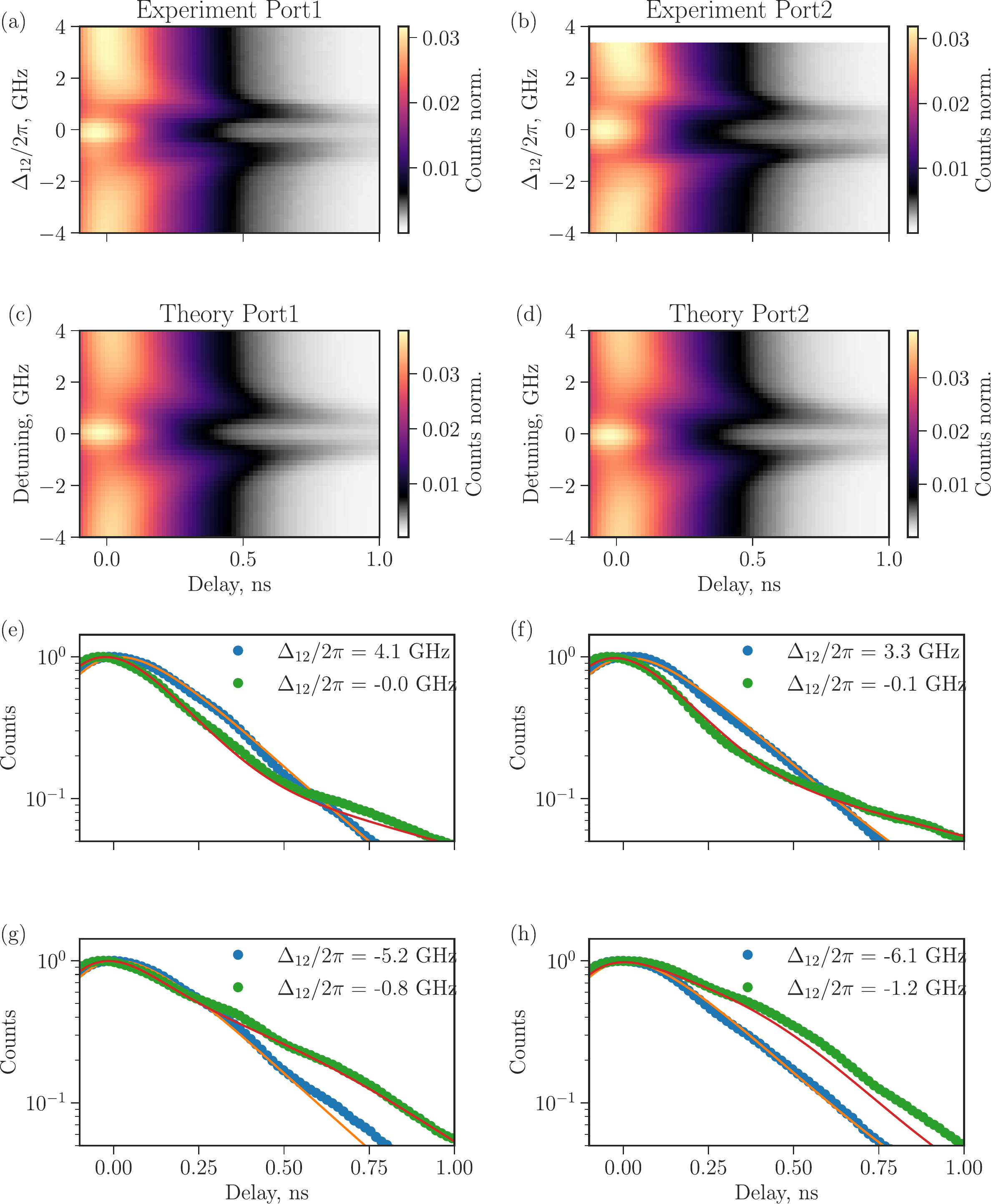}
	\end{center}
	\caption{ 
	(color online) 'Off-resonant' excitation of QD2 for different detunings from QD1 and at a magnetic field of 2.1 T. 
	(a-b) Experimental decay dynamics of  QD2 as a function of the detuning between QD1 and QD2  $\Delta_{12}$ for photon collected from port1 (a) and  port2 (b). 
	(c-d) Theoretical calculation is represented by the developed theory model.
	(e-h) Decay dynamics for different frequency detunings between QD1 and QD2  $\Delta_{12}$. For the off-resonant condition, the decay is fitted using two decay processes, where the 2nd decay rate is equal to 0.07(1) GHz, associated with a spin-flip process. The on-resonance decay consists of two decay processes corresponding to the super- and subradiant decays. The decay with  detuning close to the linewidth (green dots) contains additional modulation on top of the decay. Parameters from \tabref{tab:1} were used for the theory curves.
}
	\label{fig:SMpshell_2QD2}
\end{figure}

\begin{figure}[h!]
\begin{center}
	\includegraphics[width=0.9\linewidth]{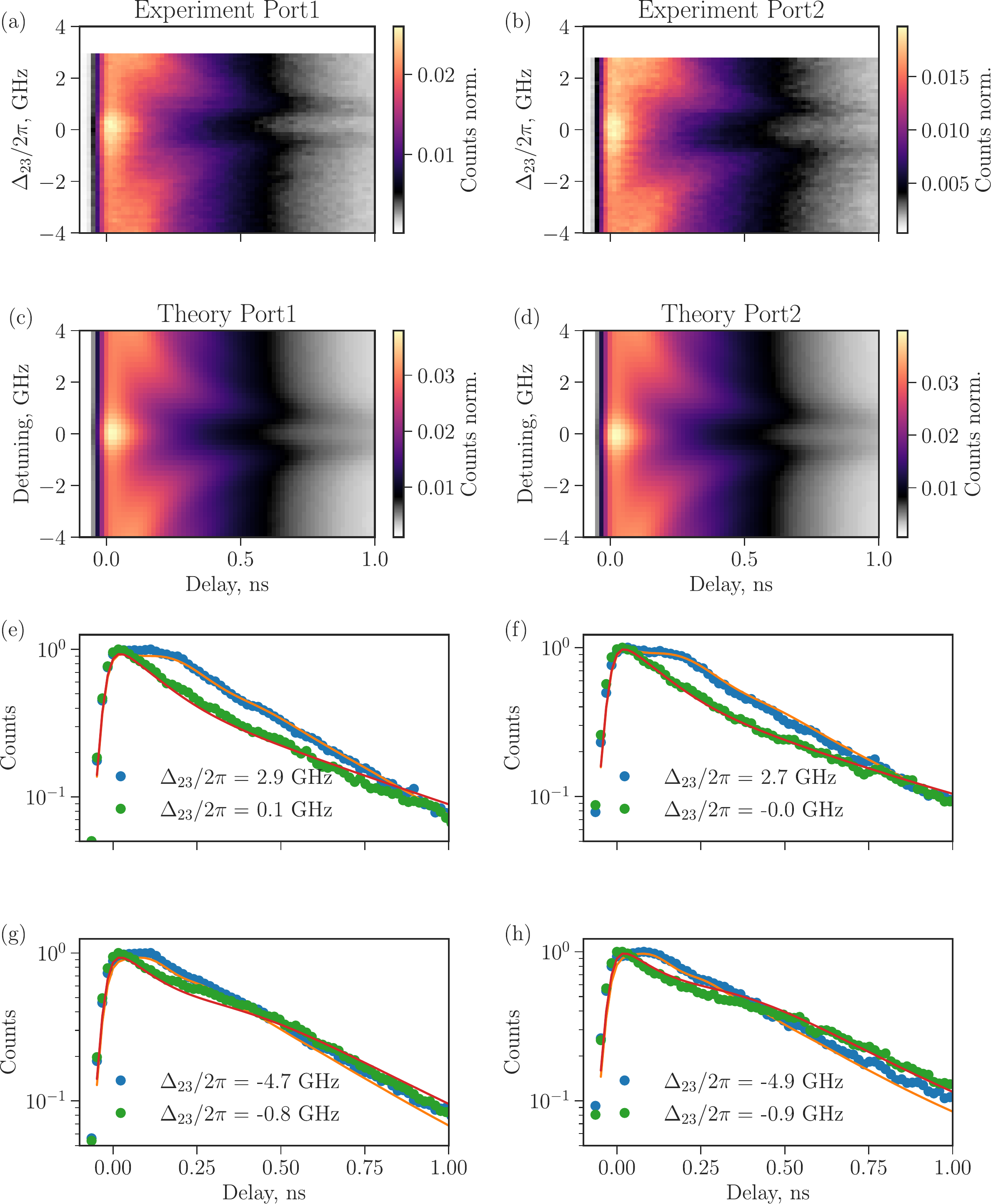}
	\end{center}
	\caption{ 
	(color online) Same as Fig. \ref{fig:SMpshell_2QD2}, but for excitation of QD3 and varying the detuning to QD2 and at a magnetic field strength of  1.0~T. 
}
	\label{fig:SMpshell_1QD3}
\end{figure}

\begin{figure}[h!]
\begin{center}
	\includegraphics[width=0.9\linewidth]{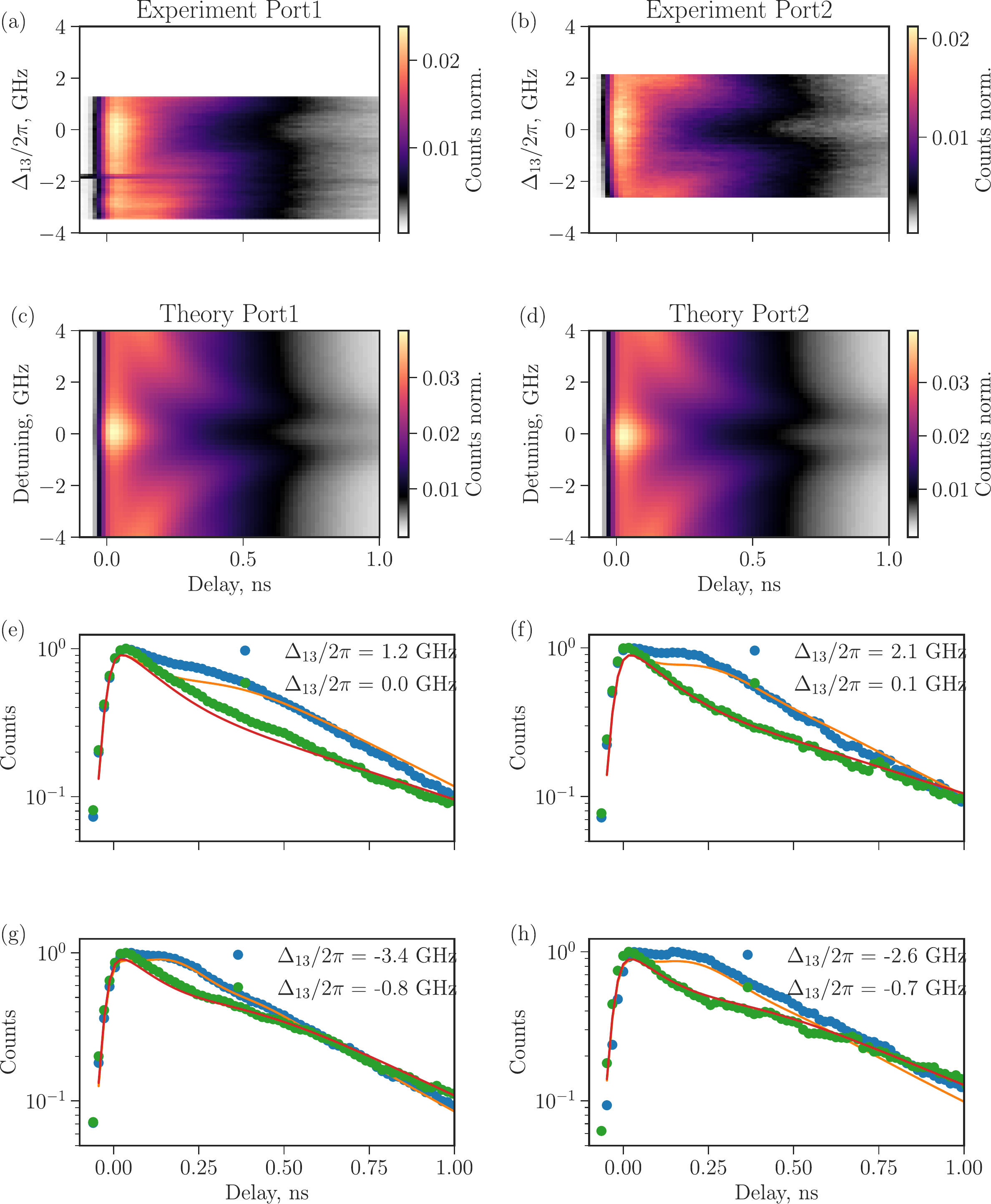}
	\end{center}
	\caption{ (color online) Same as Fig. \ref{fig:SMpshell_2QD2}, but for excitation of QD1 and varying the detuning to QD3 and at a magnetic field strength of  3.5~T. 
}
	\label{fig:SMpshell_3QD1}
\end{figure}

Following the pulsed laser excitation, the collected intensity was filtered using an optical grating setup. The intensity was collected from different ports of the PCW (\figref{fig:SMSEM}). The lifetime measurement was performed as a function of the external magnetic field  around each QD crossing. \figref{fig:SMpshell_2QD2}, \figref{fig:SMpshell_1QD3} and \figref{fig:SMpshell_3QD1} show the results of the lifetimes measurements together with the theoretical model, which uses the extracted parameters from \tabref{tab:1}. The time trace for each magnetic field was normalized to the sum of the total counts in the trace.

The detector response function measured independently was used to fit all the datasets. For QD2-QD3 crossing (\figref{fig:SMpshell_2QD2}) the SNSPD with a 200~ps FWHM temporal jitter was used. While for QD1-QD2 and QD1-QD3 crossings (\figref{fig:SMpshell_1QD3},  \figref{fig:SMpshell_3QD1}), it was a fast APD with 40~ps FWHM temporal jitter.

To find one consistent set of parameters, we use the theoretical model described above. Modeling of the data was done by using Eqs. \eqref{eq:OutputfieldsRight}, \eqref{eq:OutputfieldsLeft} to calculate the intensity in each of the ports. Spectral diffusion was taken into account by integrating the overall intensity over the spectral linewidth for detunings $\Delta$ between each pair of QDs. Each of the time traces was normalized to the sum of the trace. For the fitting, five time traces were used at different detunings around zero detuning. The results of the fit are shown in \tabref{tab:1}. 
$\sqrt{\beta_{2}\beta_{3}} = 0.8$ is used during the fitting. This is a conservative value leading to a lower bound on the coherent coupling constants that are extracted from the fit.

The extracted coupling constants $\Gamma_{ij}$ and $J_{ij}$ are very similar for all three QD pairs leading to a similar accumulated phase lag $\phi_{ij}$ corresponding to predominantly dissipative coupling. This can be explained by the fact that all three QDs were selected using the resonant transmission measurement by giving strong extinction in resonant transmission measurements (\figref{fig:SMRTfit}). With this approach, we pre-select only the QDs strongly coupled to the mode of the PCW (i.e., with a high $\beta$-factor). Since the $\beta$-factor has a strong spatial dependence, this condition results in having QDs at well-defined positions inside the unit cell of the PCW \cite{Javadi2018}. 

Close to the band edge of the PCW, the wavevector is given by $k a = \pi$, where $a$ is the lattice constant of the photonic crystal. The phase lag between two QDs separated by the integer number of unit cells $x_{ij} = N a$  is then $\phi_{ij} = N\pi$, corresponding to dissipative coupling. As explained in Section~\ref{sec:SMtheory}, the effects of directional chiral coupling can be ignored. The effects from  local phases due to QDs positioned off-axis in the PCW can be incorporated into the definition of the phase $\phi$ while keeping the same definition for the coupling constants Eq.~\eqref{eq:GJ}.

We notice that the separations between QD pairs  are very close to an integer number of unit cells of the photonic crystal with period $a = 240$~nm: $x_{12} = 5.2(1) a$,  $x_{23} = 4.0(1) a$ and  $x_{13} = 9.2(2) a$, respectively. This leads to  similar contributions from dissipative $\Gamma_{ij}$ and dispersive $J_{ij}$ coupling for different QD pairs. 

\label{sec:SMpshell}

\label{sec:SMSaturationpshell}

\clearpage
\newpage

\thispagestyle{empty}

\clearpage
\newpage

\section{Resonant excitation}
\begin{figure}[!hb]
\begin{center}
\includegraphics[width=0.75\linewidth]{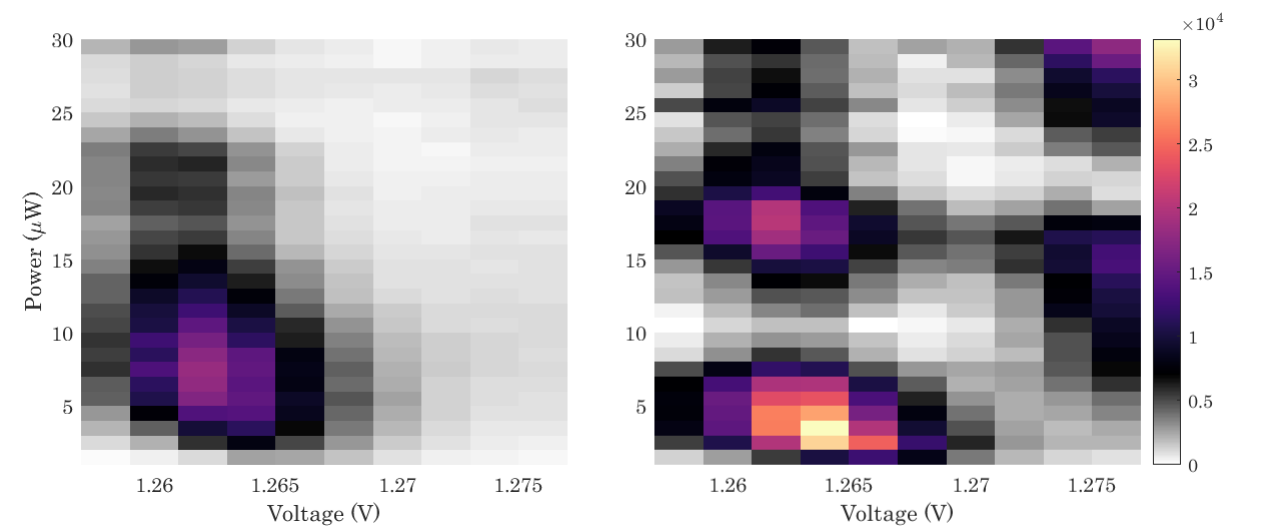}
\end{center}
\caption{\textbf{Rabi oscillations.} Left: Exciting only one dipole (QD3) of the QD2-QD3 pair. Right: Exciting both dipoles after rotating the polarization of the excitation laser. The dipole of QD2 is resonant with the filtering etalon when applying a voltage of around 1.277~V. Maxima appear for two different values of the bias voltage, indicating that both dipoles are populated for different powers. The measurement is done at the magnetic field of 0.8 T, using a 5 ps laser pulse centered around 940.5 nm.
}
\label{fig:SMRabi}
\end{figure}

The dynamics shown in \figref{fig:2}(e),(f), and \figref{fig:3}(e) in the main text are induced by resonant (s-shell) excitation. In this case, the pulsed laser frequency was tuned into resonance with the emission frequency of the driven QDs, i.e., about $318.76$~THz. The laser was suppressed in the collection path with both a transmission grating and a temperature-controlled etalon filter, where the latter has a narrow linewidth of $3$~GHz. The excitation pulse from the laser was stabilized with an AOM and frequency filtered.  The power for the laser pulse was calibrated based on Rabi oscillations with the QDs detuned from each other (\figref{fig:SMRabi}). 

\begin{figure}[h!]
\begin{center}
	\includegraphics[width=0.9\linewidth, trim=0.35cm 0.55cm 0.35cm 0.4cm,clip]{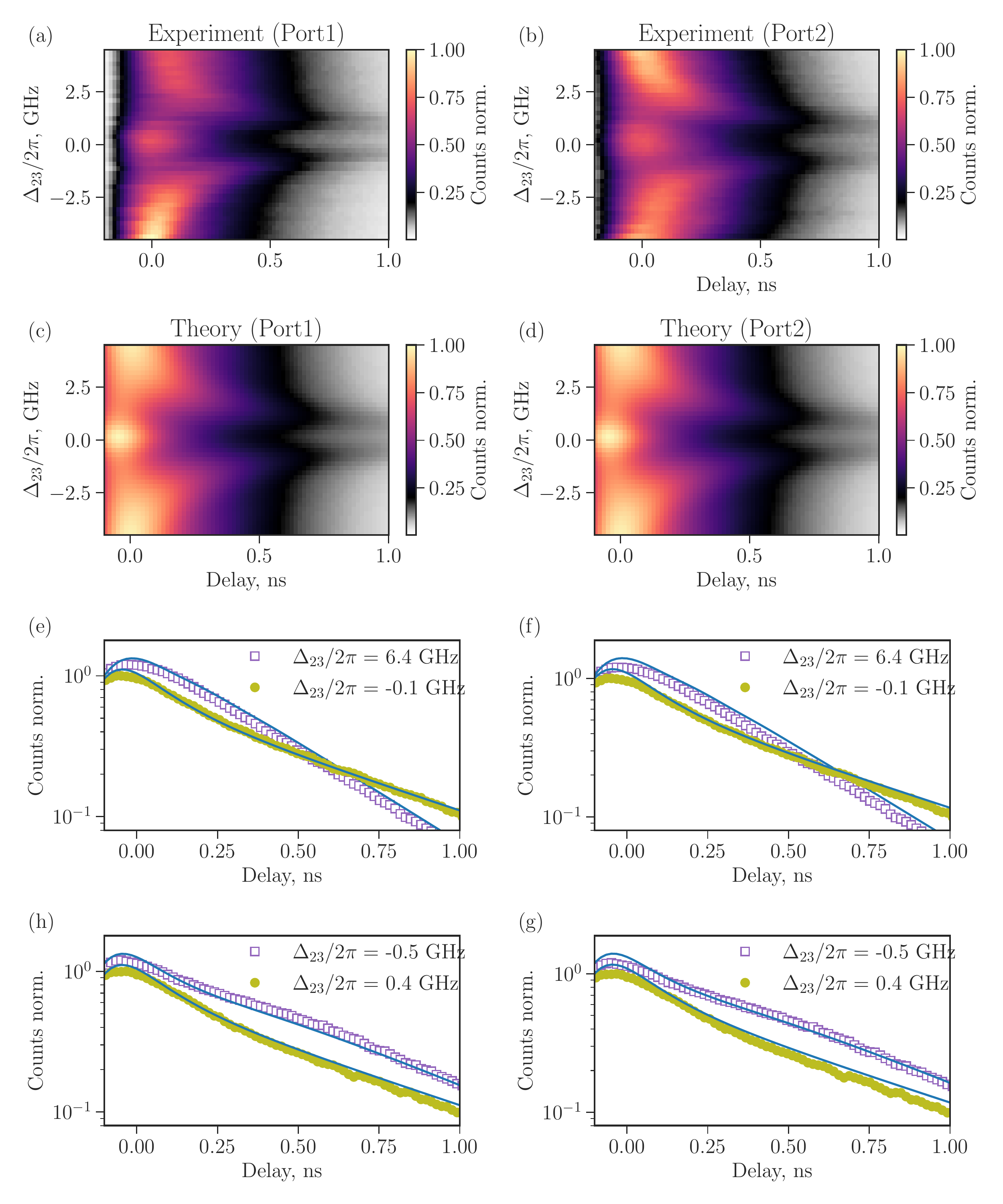}
	\end{center}
	\caption{ 
	(color online) Resonant excitation of QD3 around the resonance with QD2 at 1.05 T. 
	(a-b) The experimental excitation dynamics of  QD3 as a function of the detuning between QD2 and QD3  $\Delta_{23}$ for photon collected from  port1 (a) and  port2 (b). 
	(c-d) Theoretical calculation represented by the developed theory model.
	(e-h) Decay dynamics for different frequency detunings between QD2 and QD3  $\Delta_{23}$.  The on-resonance decay consists of two decay processes corresponding to the super- and subradiant decays. For the detuned case, an additional  coherent modulation is superposed on top of the decay. Parameters from \tabref{tab:1} were used for theory curves.
}
	\label{fig:SMpshell_1QD3_RF}
\end{figure}

Results of the excitation of QD3 when tuned close to resonance with QD2 are shown in \figref{fig:SMpshell_1QD3_RF}. The recorded behavior is equivalent to the measurement results with 'off-resonant' excitation. 

\begin{figure}[h!]
\begin{center}
	\includegraphics[width=0.85\linewidth, trim=0.35cm 0.55cm 0.35cm 0.4cm,clip]{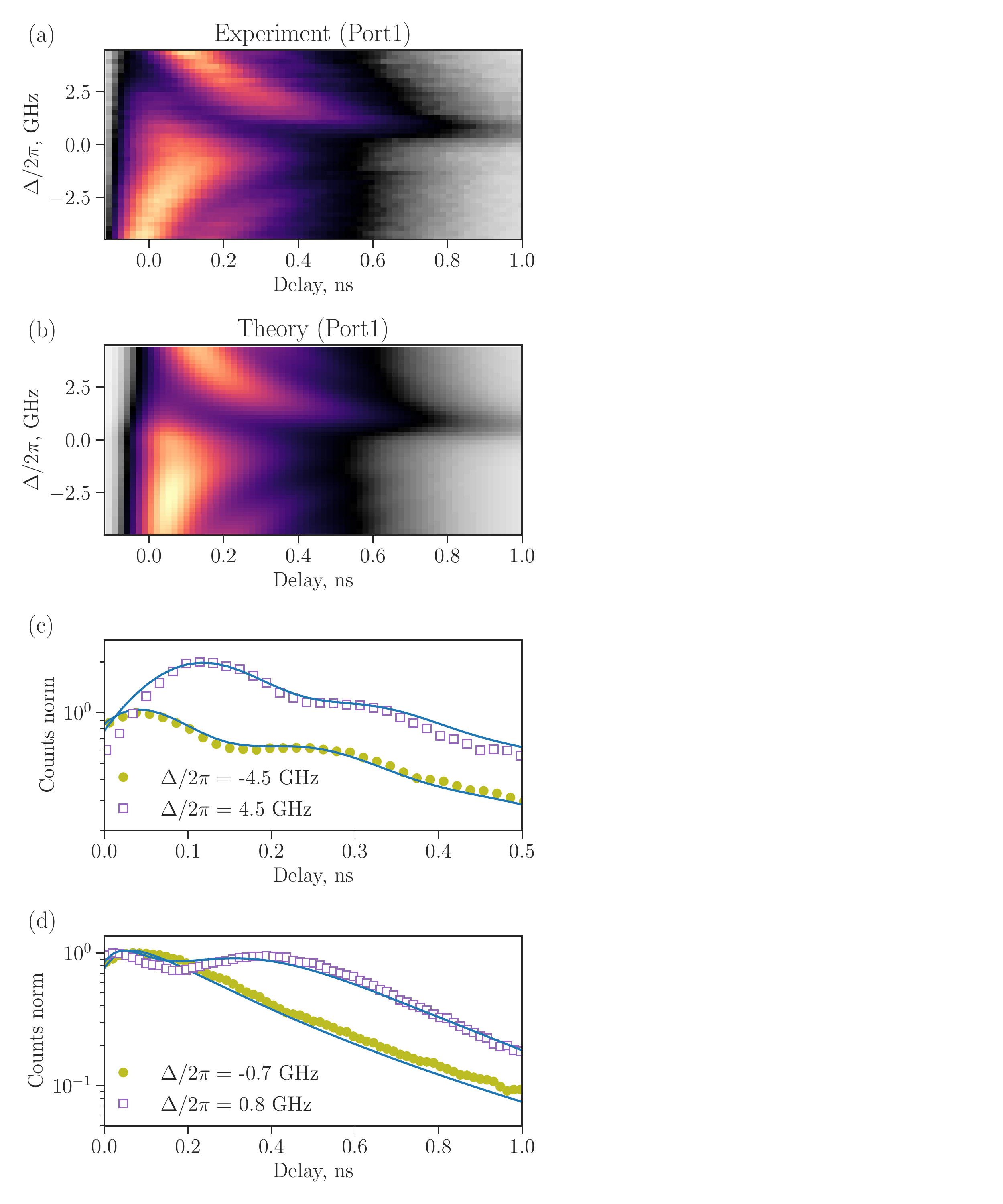}
	\end{center}
	\caption{ 
	(color online) Resonant excitation of the QD2-QD3 pair around their resonance at 1.05 T. 
	(a) Experimental excitation dynamics of the QD2-QD3 pair as a function of the detuning between QD2 and QD3  $\Delta_{23}$ for photons collected from the port1. 
	(b)Theoretical calculation represented by the developed theory model.
	(c-d) Decay dynamics for different frequency detunings between QD2 and QD3  $\Delta_{23}$. Parameters from \tabref{tab:qd2qd3} were used for theory curves.
}
	\label{fig:SMpshell_1QD23_RF}
\end{figure}

The difference becomes clear when exciting both QDs from the pair, see \figref{fig:SMpshell_1QD23_RF}. For these measurements, the same setup is used while the polarization of the driving laser field is adjusted to pump both QDs. The phase difference between two excitation fields is $\theta \approx \pi/2$, which is set by the local polarization projection of the driving field to two dipoles of the crossing. The two QD transition dipole moments are orthogonally (circularly) polarized in the Faraday magnetic field configuration, see \figref{fig:SMpcw}a., corresponding to the transitions $\ket{0} \longleftrightarrow \ket{\uparrow\Downarrow}$ and $\ket{0} \longleftrightarrow \ket{\downarrow\Uparrow}$, respectively. The polarization control of the driving field allows to effectively control the phase $\theta$ between two driving fields.

The fitting of the phase $\theta$ and driving field areas $\Omega_1$  and $\Omega_2$ was done using the theoretical model described above. The parameters extracted from the previous fit \tabref{tab:1} for the QD2-QD3 pair were used, while fitting only $\theta$, $\Omega_1$ and $\Omega_2$ parameters. The results are given in \tabref{tab:qd2qd3}.

\begin{table}
\caption{\textbf{Parameters of the photon-mediated coupling for QD2-QD3 pair while driving  both QDs}: radiative recombination rates $\{\Gamma_{2}, \Gamma_{3}\}$,  spectral diffusion linewidth $\sigma_{sd}$, phonon dephasing rate $\gamma_d$, phase lag $\phi_{23}$, dissipative coupling rate $\Gamma_{23}$ and dispersive coupling rate $J_{23}$, {$\Omega_2$, $\Omega_3$} excitation pulse areas, $\theta$ phase between two driving fields.}
\label{tab:qd2qd3}
\setlength{\tabcolsep}{5pt}
\renewcommand{\arraystretch}{1.4}

\begin{center}
\begin{tabular}{c|c}
\toprule[1pt]\midrule[0.3pt]
$\{\Gamma_{2}, \Gamma_{3}\}/2\pi$, GHz & 0.79, 0.73   \\
$\sigma_{\text{sd}}/2\pi$, GHz & 0.38   \\ 
$\gamma_d/2\pi$, GHz & 0.03  \\  
$\phi_{23}$, rad &  0.05      \\ 
$\Gamma_{23}/2\pi$, GHz &  0.61       \\  
$J_{23}/2\pi$, GHz &  0.03      \\  
$\Omega_2$, $\Omega_3$ &  0.87(6), 1.33(5)      \\ 
$\theta$ & $-0.48(2)\pi$  \\ 
\midrule[0.3pt]\bottomrule[1pt]  
\end{tabular}
\end{center}
\end{table}

\end{document}